\documentclass[2p]{elsarticle}

\usepackage{fancyhdr}
\usepackage{ctable}
\usepackage{amsmath}
\usepackage{mathtools}
\usepackage{commath}
\usepackage{graphicx}
\usepackage{textcomp} 

\usepackage{graphics}
\usepackage{tabularx}
\usepackage{color}
\usepackage{geometry}
\usepackage{natbib}
\usepackage{mciteplus}
\usepackage{footnote}
\usepackage{amssymb}
\usepackage{pdflscape}
\usepackage{afterpage}
\usepackage{multirow}
\usepackage[title]{appendix}
\usepackage{subfigure}
\usepackage{caption}
\usepackage{lineno}
\usepackage{amssymb}
\usepackage{natbib}
\usepackage{mleftright} 
\usepackage{latexsym}
\usepackage{color,soul}
\usepackage[intoc]{nomencl}
\usepackage{threeparttable}
\usepackage{booktabs} 
\usepackage{nomencl}  
\usepackage{longtable}   
\usepackage{array}       

\makenomenclature     
\usepackage{array}
\newcolumntype{P}[1]{>{\centering\arraybackslash}p{#1}}
\makenomenclature
 
\usepackage{etoolbox}
\renewcommand\nomgroup[1]{%
	\item[\bfseries
	\ifstrequal{#1}{A}{}{%
		\ifstrequal{#1}{G}{Greek Letters}{%
			\ifstrequal{#1}{S}{Subscripts}{}}}%
	]}

\usepackage[version=3]{mhchem} 

\biboptions{numbers,sort&compress}

\begin{document}
\begin{frontmatter}

\title {Impacts of flow velocity and microbubbles on water flushing in a horizontal pipeline}

\author [First]{Mohammadhossein Golchin}
\author [First] {Siyu Chen}
\author[First]{Shubham Sharma}
\author [Second]{Yuqing Feng}
\author[Third]{George  Shou}
\author[First]{Petr Nikrityuk}
\author[Fourth]{Somasekhara Goud Sontti\corref{cor2}}
\ead{somasekhar.sonti@iitdh.ac.in}
\author[First]{Xuehua Zhang\corref{cor1}}
\ead{xuehua.zhang@ualberta.ca}

\address[First]{Department of Chemical and Materials Engineering, University of Alberta, Alberta T6G 1H9, Canada}

\address[Second]{Mineral Resources Business Unit, CSIRO, Clayton, VIC 3169, Australia}

\address[Third]{BRASS Engineering International, 2551 San Ramon Valley Blvd \#226, San Ramon, CA 94583, United States}

\address[Fourth]{Multiphase Flow and Microfluidics (MFM) Laboratory, Department of Chemical Engineering, Indian Institute of Technology
	Dharwad, Dharwad, Karnataka 580011, India}

\cortext[cor1]{Corresponding author}
\cortext[cor2]{Corresponding author}

\begin{abstract}
\noindent 

Water flushing to remove particle sediment is essential for safe and continuous transport of many industrial slurries through pipelines. An efficient flushing strategy may reduce water consumption and the cost associated with water usage, and help water conservation for sustainability. In this study, a computational fluid dynamics (CFD) model coupled with the kinetic theory of granular flow for the flushing process is presented. The CFD models were validated against field data collected from a coal slurry pipeline of 128 $km$ in length, 0.575~$m$ in diameter, achieving an average error of less than 15\% for outlet solid concentration over time. A parametric study evaluated the effects of water velocity (1.88-5.88~$m/s$), bubble size (50~$\mu m$, 150~$\mu m$, and 1000~$\mu m$) and bubble volume fraction (0.05-0.2) on flushing performance including pipeline cleanness, cleanness efficiency, and water consumption. The obtained outcomes indicate that higher water velocity is preferred and an increase in water velocity from $1.88~m/s$ to $5.88~m/s$ reduces the water consumption by $28\%$. Large bubbles may hinder the flushing process and increase the water consumption by $23\%$. Remarkably, small bubbles facilitates the flushing process and lead to $35\%$ reduction in water consumption.  These effects are attributed to the turbulent characteristics in the pipelines in presence of microbubbles.
\end{abstract}


\end{frontmatter}

\newpage
\section{Introduction}

Hydraulic pipeline transport is an energy-efficient method for long-distance conveyance of slurries, with critical applications in mining, energy production, and wastewater management \cite{intro1pullum,intro2Mohaibes,intro3li}. However, solid particle deposition in the pipe reduces pipeline capacity, and risk blockages or even pipe damage \citep{gilles1,hosseini2024coupled,blockadeintrolep,blockade2scott}. 
Water flushing, a process where clean water scours accumulated sediments, is an effective approach for maintaining safe and uninterrupted operation of slurry pipeline \citep{flushingprocess}. As the flushing process consumes a large volume of clean water, optimizing the flushing process remains vital to minimize operational downtime, energy consumption, and water usage. Understanding of transient flow behavior during water flushing is the foundation for designing effective flushing processes.  

The flushing process is complicated, requiring sufficient energy to ensure all sediment is removed. The success of flushing hydrotransport pipeline of particle deposit depends on the properties of the deposit and water velocity. In general, the onset of sediment removal by flushing fluid is triggered by the upstream-downstream pressure gradient \cite{li2022multi}.  The critical velocity refers to the level at which the motion of a particle initiates. For quick estimations and preliminary analysis, the Shields parameter is a fundamental parameter in sediment transport \cite{camenen2008general}, providing insights into the critical conditions required for sediment transport initiation and the stability of the sediment. The critical Shields parameter depends on factors such as sediment composition, shape, size distribution, and flow conditions. However, these semi-empirical equations are limited to simple pipe geometry, and homogenous particles under simple flow conditions, not suitable for prediction of flushing heterogeneous, multiphase and multicomponent slurries, due to the complicated slurry composition and highly turbulent nature of the flow.

Computational fluid dynamics (CFD) simulations offer a powerful method to analyze and predict sediment transport in complex systems and dealing with intricate flow patterns. High-fidelity validated models can enable parametric studies and support design optimization. Based on the literature on numerical models \cite{DEM_mono2023,CFDCGDEM,zhao2022review}, generally two methods are developed to simulate slurry flow: the Eulerian-Lagrangian (E-L) method and the Eulerian-Eulerian (E-E) method. The E-L method describes the movement of each single particle using Newton's law of motion such that it tracks the trajectories of each particle and the interactions between particles such as collisions. However, it is computationally expensive and is only suitable to model system with dilute particles, such as scouring process around a monopile or pipes \cite{DEM_mono2023, CFDCGDEM}. The Eulerian-Eulerian (E-E) method describes different phases as interpenetrating continua, and applies governing equations to each of the phases while considering interactions between phases as interphase forces. Due to the lower computational complexity, the E-E model is widely used to study flows involving slurries \cite{ofei2016eulerian,thakur2022hydrodynamic}. The mixture model is a simplified E-E model that further reduces the computational time. When the interactions between secondary phases are negligible, the governing equations for different phases can be reduced to a single equation that describes the overall behavior of the multiphase flow, with a term known as drag forces that account for relative motions between phases. The mixture model is preferably used to model poly-dispersed flows such as slurry flow in a pipeline~\cite{messa2021computational}.

By using two-dimensional (2D) models, \citet{E-EI2} studied the flushing process in a lab-scale bended pipeline using a steady state model and showed that the solid content is more preferable to remain in the bended region and it requires high velocity flow to remove these sediments. \citet{sewer_DEM2} studied the role of waves on removal of deposit in flushing the sewerage using an unsteady model that couples CFD and discrete element method. However, 2D model can only capture highly symmetrical flow field and it neglects the three-dimensional (3D) characteristics of the turbulence and the interaction between solid bed and wall.  To simulate the flow field during slurry transportation in a pipeline using 3D models,  \citet{swirl-shi} investigated the transportation of multi-sized slurry flows in horizontal pipelines under swirling flow. They found that the increase in swirl level helped to form more uniform slurry flow. \citet{januario2020cfd} studied the critical velocity of water flow required to suspend dilute solid particles of different sizes in a pipeline.

Recently, a new cleaning technology for water supply pipeline has drawn attention where a binary-phase flow of air and water is used for flushing pipes with a large diameter or a complicated geometry \cite{airbubbleflush,bai2024effect}. During air-water flushing, bubble dynamics can create the fluctuation of the pressure and velocity in the pipe and a large pressure gradient to drive the particle movement. In addition, the gas in the flow reduces the head loss in a pipe with a large diameter. The bubbly water was found to achieve the same flushing effect with less water that is consumed for single-phase clean water in flushing water supply pipeline \cite{airbubbleflush}. Bubbles in a slurry flow can be generated in several methods, for example, by incorporating a Venturi tube in the pipeline \cite{motamed2020microbubble,gao2021formation,zhou2022microbubble,zhou2025experimental}. The application of bubbly water for flushing slurry may have the potential to greatly reduce water usage. 

Recent studies reported both mixture and E-E models to simulate the turbulent, bubbly, multiphase slurry flow in an industrial-scale horizontal pipeline \cite{E-EI9,SCFD,Sekhar2023POF}. The simulation results of the flow profile were in good agreement with the profiles in the field data. The size and fraction of bubbles on solid distribution were determined under various flow conditions. Population balance model was integrated in the E-E models to describe the coalescence and breakage of bubbles \cite{sontti2023numerical}. Despite the importance of the flushing process, no predictive models exist for evaluating water consumption in large-diameter pipelines, particularly for systems with multiple solid phases alongside with microbubbles. The effects of key factors such as flow velocity and bubble size and fraction on the multiphase flow remain underexplored. 

This study will present a three-dimensional (3D) numerical model to investigate the flushing process in an industrial scale pipeline. The model was validated against multiple sets of field data, focusing on solid concentration at the outlet intersection over time. A comprehensive parametric study evaluated the effects of flushing velocity, particle size, the size and fraction of bubbles on flushing efficiency and water consumption. 
The models developed in this work can be applied to identify the water system required for flushing pipeline with bubbles. Efficient flushing strategy may reduce water consumption and the cost associated with water usage, and help water conservation. The models developed in this work is also transferrable to other industrial and environmental processes of slurry transport in pipeline. 


\section{Industrial flushing process in the pipeline of coal slurry}

\begin{figure*}[!ht] 
	\centering
	\includegraphics[width=1\columnwidth]{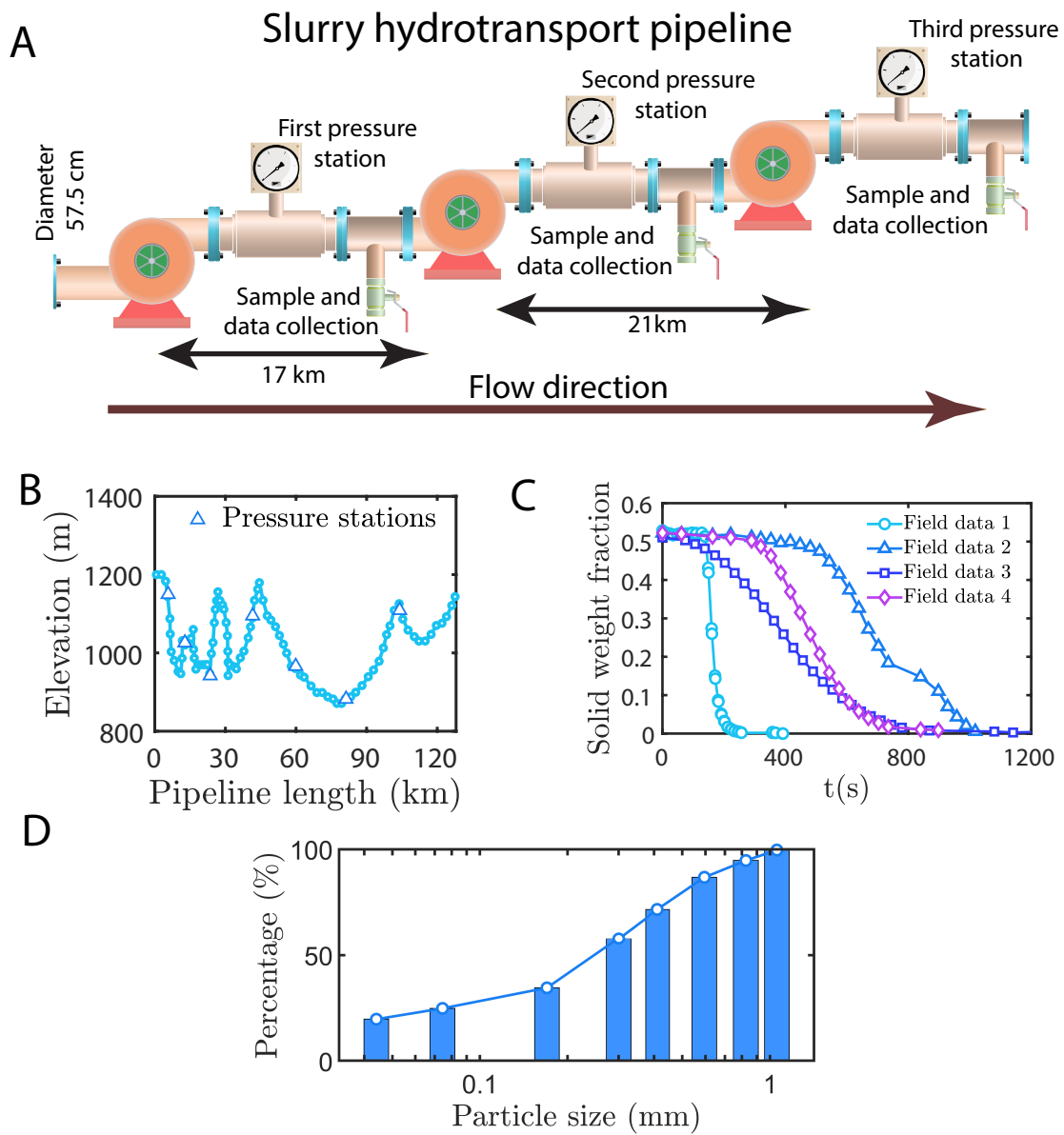} 
	\renewcommand{\captionfont}{\footnotesize}
	\caption{(A) Schematic representation of the coal slurry hydrotransport pipeline, (B) pipeline elevation through 128 $km$ distance, (C) solid content weight fraction from density meter, and (D) particle size distribution of solid content.}
	\label{fdata}
\end{figure*}

Figure \ref{fdata} A illustrates the schematic of a horizontal industrial pipeline system used to collect field data for validating a CFD model. The pipeline spans 128 kilometers in China. The internal diameter of the pipe is 57.5 centimeters. The route crosses terrain with elevations ranging from 800 to 1300 meters above sea level, as illustrated in Figure \ref{fdata} B. Along the pipeline, seven pressure and monitoring stations are positioned to track the flow condition and collect field data, including solid content (weight fraction) and particle size distribution (PSD). The four stations are located at Shenmu (Pressure station 1) with the elevation 1150 m and distance of 6 km from inlet, Jiaxian(Pressure station 2) with the elevation 1100 m and distance of 44 km from inlet, Qingjian (Pressure station 3) with the elevation 890 m and distance of 83 km from inlet, and Jiangtou (Pressure station 4) with the elevation 1120 m and distance of 105 km from inlet. These stations are strategically distributed across the pipeline elevation profile, enabling analysis of how solids behavior changes along the flow path.

An inline, non-invasive density meter was installed near the pipeline outlet 50 $m$ from each pump station, providing continuous measurements of mixture density and real-time estimation of solid content by weight fraction. The solid content weight fraction varied significantly over time during the flushing process, ranging from as high as 53\% (50\% coal particles and 3\% sand particles) down to 0\%, as demonstrated in Figure \ref{fdata} C. The solid content reduction reflects the dynamic nature of the flushing process. Simultaneously, Flow rate data were also recorded in sync with the density readings and were recorded every 10 seconds and remained relatively constant at 1755 $m^3/hr$, with a fluctuation range of approximately $\pm$60 $m^3/hr$ These synchronized measurements produced a comprehensive dataset of approximately 100,000 data points over the monitoring period.\\

Mixture samples were also collected by a sample collector manually every 12 hours from the pressure stations. The composition was PSD was analyzed after drying the slurry to separate water and solids, while PSD was determined through sieving techniques, the PSD of solid content in the pipeline is illustrated in Figure \ref{fdata} D. The range of particles in the cumulative percentage of the coarse particles is from 0.04 mm to 1.05 mm. The major coarse solid components containing more than 90\%  of the slurry are in the range of 0.15 mm to 0.7 mm. Table \ref{tab:PSD} presents the slurry PSD during the transportation stage. After the slurry transportation phase, flushing will take place, which will remove the solid content with the same PSD, which is referred to as $PSD_{1}$. The volume fraction of the solid content in the slurry for 53 weight percent is also represented in Table \ref{tab:validationCases}.\\

\begin{table}[!ht]
	\renewcommand\thetable{1}

	\centering
	
	\caption{Particle size distribution in slurry.}
	\label{tab:PSD}
	\renewcommand{\arraystretch}{1}
	\begin{centering}
		\renewcommand{\arraystretch}{1}  
		\setlength{\tabcolsep}{4pt}  
		
		\begin{tabular}{l l l l l }
			
			\hline
			phase & Density ($kg/m^3$)& Diameter ($\mu m$) & Percentage(\%) & Vf(*) in slurry \\
			\hline
			Sand & 2300 & 218  & 2.48 & 0.011 \\
			Sand & 2300 & 547 & 2.94 & 0.013 \\
			Coal & 1420 & 161 & 5.20 & 0.023 \\
			Coal & 1420 & 290 & 24.89 & 0.11 \\
			Coal & 1420 & 411 & 47.51 & 0.21 \\
			Coal & 1420 & 672 & 16.98 & 0.075 \\
			\hline
			(*) Vf : & Volume fraction & & & \\
			\hline
			
		\end{tabular}
	\end{centering}
\end{table}

To validate the CFD simulation, four flushing processes were selected at different time points during the operation of the transportation and flushing process from the in-situ field data. In each selected case, the water flow during a single flushing operation passes through each pressure station. The data from this operation was selected for validation, and average values of flow rate and the solid content volume fraction during this operation were used as input parameters for simulation. The four flushing cases are presented in Table \ref{tab:validationCases}, indicating the slurry transport and flushing process for each validation case. All the validation cases solid volume fraction conversion from weight fraction to volume fraction conversion are presented in Table S1.  \\

\begin{table}[!ht]
	\renewcommand\thetable{2}
	\caption{CFD validation cases.}
	\label{tab:validationCases}
	
	\begin{threeparttable}
		\centering
		\renewcommand{\arraystretch}{1.05}  
		\setlength{\tabcolsep}{4pt}         
		
		\resizebox{\textwidth}{!}{%
			\begin{tabular}{ 
					>{\raggedright\arraybackslash}p{3.2cm} 
					| >{\centering\arraybackslash}p{1.2cm} 
					| >{\centering\arraybackslash}p{2.7cm} 
					| >{\centering\arraybackslash}p{2.0cm} 
					| >{\centering\arraybackslash}p{1.8cm} 
					| >{\centering\arraybackslash}p{2.3cm} 
				}
				\hline
				\textbf{Validation\newline Case No.} 
				& \textbf{Ps($\ast$)\newline No.} 
				& \textbf{P = Operation} 
				& \textbf{Inlet swf($\ast\ast$)} 
				& \textbf{PSD} 
				& \textbf{Inlet velocity\newline (m/s)} \\
				\hline
				Validation Case 1 & 1 & slurry transport & 0.53  & $PSD_{1}$ & 1.88 \\
				& 1 & flushing process & 0     & $PSD_{1}$ & 1.88 \\
				Validation Case 2 & 2 & slurry transport & 0.527 & $PSD_{1}$ & 1.97 \\
				& 2 & flushing process & 0     & $PSD_{1}$ & 1.97 \\
				Validation Case 3 & 3 & slurry transport & 0.525 & $PSD_{1}$ & 1.74 \\
				& 3 & flushing process & 0     & $PSD_{1}$ & 1.74 \\
				Validation Case 4 & 4 & slurry transport & 0.520 & $PSD_{1}$ & 2.23 \\
				& 4 & flushing process & 0     & $PSD_{1}$ & 2.23 \\
				\hline
			\end{tabular}
		} 
		
		\vspace{2pt}
		\begin{tablenotes}
			\small
			\item ($\ast$) Ps: Pressure station \quad ($\ast\ast$) swf: solid weight fraction
		\end{tablenotes}
	\end{threeparttable}
\end{table}

	

\section{CFD model implementation}

\begin{figure*}[!ht] 
	\centering
	\includegraphics[width=1\columnwidth]{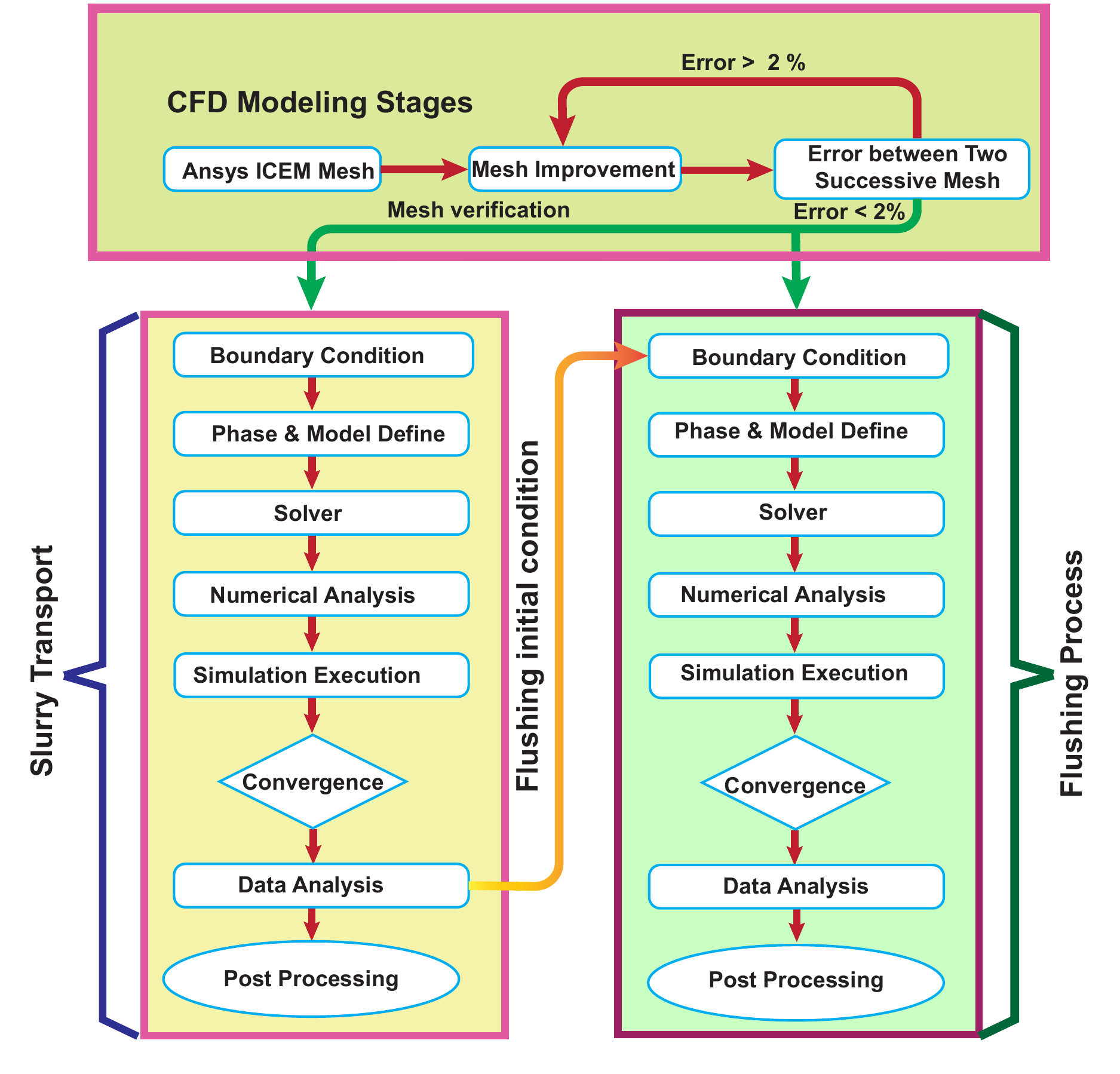} 
	\captionsetup{font={normal}}
	\renewcommand{\captionfont}{\footnotesize}
	\caption{CFD modeling flowchart, detailing mesh generation and verification, followed by sequential CFD simulation stages for slurry transport and flushing processes, including convergence and post processing sections.}
	\label{f077}
\end{figure*}

Figure \ref{f077} outlines the steps involved in the numerical model for the process. The slurry flushing modeling is performed in two stages: slurry transport and flushing process. To prepare cases for CFD flushing simulation, slurry is first introduced into the pipeline to achieve a steady-state distribution and formation of a solid bed. After the slurry has been fed, the pipeline contains deposits of particles of various sizes, which serve as the initial condition for the flushing process. Clean water is then introduced to the pipeline to simulate the flushing operation, during which the solid bed is removed.





CFD methodology in Figure \ref{f077}, the simulation consists of two sequential stages. These stages represent two different modeling objectives, and the time required for each stage depends on the underlying physical process. The flushing process (i.e., removal of the solid content in the pipeline) typically takes longer to completely dislodge and transport the settled solids, which necessitated a longer simulation time (i.e., more time steps). The parametric study of the flushing process examines a different range of operating conditions. Table \ref{tab:parameterrange} presents the range of simulation parameters used in the parametric study during flushing process, which aims to remove residual solids following slurry transport with a solid content of 53 wt\%. The parametric study focuses on the effects of various parameters such as velocity (Cases 1, 2, 3, 4, and 5), Bubble size (Cases 1, 6, 7 and 8), and bubble volume fraction with a fixed size of 1000 $\mu m$ (Cases 1, 8, 9 and 10).

\begin{table*}[!ht]
	\centering
	\renewcommand\thetable{3}
	\caption{Simulation parameters configuration.}
	\label{tab:parameterrange}
	\begin{center}
		\renewcommand{\arraystretch}{1.1}  
		\setlength{\tabcolsep}{4pt}  
		\begin{tabular}{l l l l l l l }
			\hline
			& &  Flushing operating & condition & & & \\
			\hline
			& Case No. & Velocity (m/s) & Bubble vf & Bubble size ($\mu$m) & PSD \\
			\hline
			Range of parameters & & & & & & \\   
			Velocity & & & & & & \\
			& Case 1 & 1.88 & 0 & ------ & $PSD_{1}$\\
			& Case 2 & 2.88 & 0 & ------ & $PSD_{1}$\\
			& Case 3 & 3.88 & 0 & ------ & $PSD_{1}$\\
			& Case 4 & 4.88 & 0 & ------ & $PSD_{1}$\\
			& Case 5 & 5.88 & 0 & ------ & $PSD_{1}$\\
			Bubble size & & & & & & \\
			(* No bubble) & Case 1 (*) & 1.88 & 0 & ------ & $PSD_{1}$\\
			& Case 6 & 1.88 & 0.05 & 50 & $PSD_{1}$\\
			& Case 7 & 1.88 & 0.05 & 150 & $PSD_{1}$\\
			& Case 8 & 1.88 & 0.05 & 1000 & $PSD_{1}$\\
			Bubble & & & & & & \\
			& Case 1 (*) & 1.88 & 0 & ------ & $PSD_{1}$\\
			& Case 8 & 1.88  & 0.05 & 1000 & $PSD_{1}$\\
			& Case 9 & 1.88  & 0.1 & 1000 & $PSD_{1}$\\
			& Case 10 & 1.88  & 0.2 & 1000 & $PSD_{1}$\\
			\hline
		\end{tabular}%
		
	\end{center}
\end{table*}

\subsection{Governing equations}

The governing equations for the mixture model are provided in Table~\ref{tab:Eq45}. Equation (1a) stands for the continuity equation where the density of the mixture is expressed as a sum of the densities of all phases weighted by their respective volume fractions in equation (1e). Equation (1b) stands for the momentum balance, where the convection forces on the left-hand side of the equation is equal to the sum of pressure gradients, shear forces due to turbulence and velocity gradient, and body forces such as gravitational field force and interaction forces among phases on the right-hand side of the equation. The secondary phase volume fraction is expressed in Equation (1c). The mixture velocity, mixture density and mixture viscosity  of the slurry system are expressed in Equation (1d)-(1f) \cite{Ansys}. The drift velocity of secondary phases and the relative velocity are expressed in Equation (1g) and (1h). The other associated quantities are reported in Equation  (1h) to (1l). The frag force function will considering for different primary and secondary phases in the Equation (1k). The energy losses due to pipeline friction are inherently captured through the momentum equations, which include wall shear stress as part of the viscous term. The no-slip boundary condition at the pipe wall ensures accurate modeling of frictional effects. The pressure drop across the domain reflects the combined effect of frictional losses and flow resistance, which is used to estimate energy losses throughout the system.

The forces due to the interaction of the primary and secondary phases consist of the net value of drag force, turbulent dispersion force, and elastic and non-elastic internal collisions. The sand and coal phases are granular materials; the granular properties are defined for these secondary phases. Equations (2a) to (2m) in Table~\ref{tab: granularsolid} calculate the shear stress, sand and coal pressure, and viscosity of these granular materials using the KTGF model. The KTGF model equation uses the granular properties of sand and coal to define them as fluids. Interactions between the carrier fluid (primary phase) and sand and coal phases (secondary phases) are governed by drag force estimation. In this study, the Gidaspow drag model~\citep{drag1, SCFD, drag3} is used to model the interaction between the carrier and sand and coal phases in the dense slurry system.

\begin{table*}[!ht]
	\centering
	
	\caption{Mixture model modeling equations~\citep{E-EI9,Ansys}.}
	\label{tab:Eq45}
	\renewcommand{\arraystretch}{1.5}
	\setlength{\tabcolsep}{12pt}
	\begin{tabular}{ll}
		\hline
		\textbf{Equation} & \textbf{Expression} \\
		\hline
		
		Continuity & 
		$\displaystyle \frac{\partial \rho_m}{\partial t} + \nabla \cdot (\rho_m \vec{v}_m) = 0 \hfill \text{(1a)}$ \\
		
		Momentum & 
		\parbox[t]{0.85\textwidth}{
			\raggedright
			$\displaystyle \frac{\partial (\rho_m \vec{v}_m)}{\partial t} + \nabla \cdot (\rho_m \vec{v}_m \vec{v}_m) = -\nabla p + \nabla \cdot \left[\mu_m \left(\nabla \vec{v}_m + (\nabla \vec{v}_m)^T\right)\right] + \rho_m \vec{g} + \vec{F} + \nabla \cdot \left( \sum_k \alpha_k \rho_k \vec{v}_{dr,k} \vec{v}_{dr,k} \right)$ \hfill (1b)
		} \\
		
		Volume fraction (solid) & 
		$\displaystyle \frac{\partial (\alpha_s \rho_s)}{\partial t} + \nabla \cdot \left( \alpha_s \rho_s \vec{v}_m + \alpha_s \rho_s \vec{v}_{dr,s} \right) = 0 \hfill \text{(1c)}$ \\
		
		Mixture velocity & 
		$\displaystyle \vec{v}_m = \frac{\sum_{k=1}^n \alpha_k \rho_k \vec{v}_k}{\rho_m} \hfill \text{(1d)}$ \\
		
		Mixture density & 
		$\displaystyle \rho_m = \sum_{k=1}^n \alpha_k \rho_k \hfill \text{(1e)}$ \\
		
		Mixture viscosity & 
		$\displaystyle \mu_m = \sum_{k=1}^n \alpha_k \mu_k \hfill \text{(1f)}$ \\
		
		Drift velocity & 
		$\displaystyle \vec{v}_{dr,k} = \vec{v}_k - \vec{v}_m \hfill \text{(1g)}$ \\
		
		Slip velocity & 
		$\displaystyle \vec{v}_{pq} = \vec{v}_p - \vec{v}_q \hfill \text{(1h)}$ \\
		
		Mass fraction & 
		$\displaystyle c_k = \frac{\alpha_k \rho_k}{\rho_m} \hfill \text{(1i)}$ \\
		
		Drift-slip relation & 
		$\displaystyle \vec{v}_{dr,p} = \vec{v}_{pq} - \sum_{k=1}^n c_k \bar{v}_{qk} \hfill \text{(1j)}$ \\
		
		Algebraic slip velocity & 
		$\displaystyle \bar{v}_{pq} = \frac{\tau_p}{f_{\text{drag}}} \cdot \frac{(\rho_p - \rho_m)}{\rho_p} \cdot \bar{a} \hfill \text{(1k)}$ \\
		
		Particle relaxation time & 
		$\displaystyle \tau_p = \frac{\rho_p d_p^2}{18 \mu_q} \hfill \text{(1l)}$ \\
		
		\hline
	\end{tabular}
\end{table*}

\begin{table*}[!ht]
	\renewcommand\thetable{5}
	\begin{center}
		\caption{Kinetic theory of granular flow equations.\cite{Sekhar2023POF,lun,gidaspowbook,gidaspow2} (\textit{i}th solid or coal phases)}
		\label{tab: granularsolid}
		
		\renewcommand{\arraystretch}{1.1}  
		\setlength{\tabcolsep}{4pt}       
		
		\resizebox{\textwidth}{!}{%
			\begin{tabular}{l l}
				\hline
				
				$q = c$ (coal), $s$ (sand) & (2a) \\
				$\lambda_{qi} = \dfrac{4}{3} \alpha_{qi}^2 \rho_{qi} d_{qi} g_{0,ii}(1+e_{jj})\left(\dfrac{\Theta_{qi}}{\pi}\right)^{1/2}$ & (2b) \\
				$g_{0,ii} = \left[1 - \left(\sum_{i=1}^2 \dfrac{\alpha_{qi}}{\alpha_{s_{\max},c_{\max}}}\right)^{1/3}\right]^{-1} + \dfrac{d_{qi}}{2} \sum_{i=1}^2 \dfrac{\alpha_{qi}}{d_{qi}}$ & (2c) \\
				$g_{0,ij} = \dfrac{d_{qi}g_{0,ii} + d_{qj}g_{0,jj}}{d_{qi}+d_{qj}}$ & (2d) \\
				$\Theta_{qi} = \dfrac{1}{3} \left\| \vec{v'}_{qi} \right\|^2$ & (2e) \\
				$0 = \left(-P_{qi} \bar{\bar{I}} + \bar{\bar{\tau}}_{qi} \right): \nabla \vec{v}_{qi} - \gamma_{\Theta_{qi}} + \phi_{l,qi}$ & (2f) \\
				$\gamma_{\Theta_{qi}} = \dfrac{12(1 - e_{ii}^2 g_{0,ii})}{d_{qi} \pi^{1/2}} \rho_{qi} \alpha_{qi}^2 \Theta_{qi}^{3/2}$ & (2g) \\
				$\phi_{l,qi} = -3 K_{l,qi} \Theta_{qi}$ & (2h) \\
				$P_{qi} = \alpha_{qi} \rho_{qi} \Theta_{qi} \left[1 + 2 \sum_{j=1}^2 \left(\dfrac{d_{qi}+d_{qj}}{2d_{qi}}\right)^3 (1+e_{ij} \alpha_{qj} g_{0,ij}) \right]$ & (2i) \\
				$\mu_{qi} = \mu_{qi,\text{col}} + \mu_{qi,\text{kin}} + \mu_{qi,\text{fr}}$ & (2j) \\
				$\mu_{qi,\text{col}} = \dfrac{4}{5} \alpha_{qi} \rho_{qi} g_{0,ii}(1+e_{ij})\left(\dfrac{\Theta_{qi}}{\pi}\right)^{1/2} \alpha_{qi}$ & (2k) \\
				$\mu_{qi,\text{kin}} = \dfrac{10 \rho_{qi} d_{qi} (\Theta_{qi} \pi)^{1/2}}{96 \alpha_{qi}(1+e_{ij}) g_{0,ii}} \left[1 + \dfrac{4}{5} g_{0,ii} \alpha_{qi} (1 + e_{ii}) \right]^2 \alpha_{qi}$ & (2l) \\
				$\mu_{qi,\text{fr}} = \dfrac{P_{qi} \sin \phi_{qi}}{2 I_{2D}^{1/2}}$ & (2m) \\
				
				\hline
			\end{tabular}
		} 
	\end{center}
\end{table*}

\subsection{Turbulence model, wall function and rheological model for slurry}

In the flushing process in the industrial pipe, Reynolds number is usually high, indicating that the flow is turbulent. As a result, it is crucial to model the turbulence behavior in the present study. While \textbf{$k-\omega$} and \textbf{$k-\epsilon$} turbulence models are widely used in modeling the dense slurry systems~\citep{SCFD,turbu2,yiyi,Sekhar2023POF}, these models only consider isotropic turbulence~\citep{turbu1,turbu2,turbu3,turbu4,turbu5}. However, in the flushing process, the interactions of the solid particles and complex liquid flows generates anisotropic turbulence. As a result, the Reynolds stress model (RSM) is used to capture the turbulence behavior in all directions~\citep{baker2019train}. The RSM model is derived from the Reynolds-averaged Navier-Stokes (RANS) equations. The governing equations for the Reynolds stress model are listed in Table~\ref{tab: Turbulence} (Equations.~3(a)-3(c)~\citep{abdulbubble,elbowiceslurry,cicepigging,launder}). The wall effect on the flow is modelled by the standard wall functions~\citep{cGauravslurrypowder,kepsilonturb,Slurryocean,Rans23d,wliuhozslurry,yplus1,yplus2}. In the present study, the $\begin{alignedat}{8} y^{+} \end{alignedat}$ is 34.2, indicating insignificant near-wall effects.

\begin{table*}[!ht]
	\renewcommand\thetable{6}
	\centering
	\caption{Reynolds stress turbulence model~\cite{turbu4}.}
	\label{tab: Turbulence}
	
	\renewcommand{\arraystretch}{1.1}
	\setlength{\tabcolsep}{4pt}
	
	\resizebox{\textwidth}{!}{%
		\begin{tabular}{l l l}
			\hline
			
			The Reynolds stress transport equation & & \\
			\multicolumn{3}{l}{
				\(
				\begin{aligned}
					\frac{\partial}{\partial t}(\rho \overline{u_{i}' u_{j}'}) + \frac{\partial}{\partial x_{k}}(\rho u_{k} \overline{u_{i}' u_{j}'}) 
					&= -\frac{\partial}{\partial x_{k}} \left(\rho \overline{u_{i}' u_{j}' u_{k}'} + \overline{p(\delta_{kj} u_{i}' + \delta_{ik} u_{j}')} \right) \\
					&\quad + \frac{\partial}{\partial x_{k}} \left(\mu \frac{\partial}{\partial x_{k}} (\overline{u_{i}' u_{j}'}) \right) \\
					&\quad - \rho \left(\overline{u_{i}' u_{k}'} \frac{\partial u_{j}}{\partial x_{k}} + \overline{u_{j}' u_{k}'} \frac{\partial u_{i}}{\partial x_{k}} \right) \\
					&\quad - \rho \beta (g_{i} \overline{u_{j}' \theta} + g_{j} \overline{u_{i}' \theta}) \\
					&\quad + \overline{p \left( \frac{\partial u_{i}'}{\partial x_{j}} + \frac{\partial u_{j}'}{\partial x_{i}} \right)} 
					- 2\mu \overline{\frac{\partial u_{j}'}{\partial x_{i}} \frac{\partial u_{j}'}{\partial x_{i}}} \\
					&\quad - 2\rho \Omega_{k} \left( \overline{u_{j}' u_{q}'} \epsilon_{ikq} + \overline{u_{i}' u_{q}'} \epsilon_{jkq} \right) + S
				\end{aligned}
				\) \hfill (3a)
			} \\
			
			\\[-1ex]
			Turbulent kinetic energy and dissipation rate & & \\
			\multicolumn{3}{l}{
				\(
				k = \frac{1}{2} \overline{u_{i}' u_{i}'}
				\) \hfill (3b)
			} \\
			
			\multicolumn{3}{l}{
				\(
				\epsilon = 2\mu \overline{\frac{\partial u_{j}'}{\partial x_{i}} \frac{\partial u_{j}'}{\partial x_{i}}}
				\) \hfill (3c)
			} \\
			
			\hline
		\end{tabular}
	} 
\end{table*}


The slurries of fine sand and coal particles suspended in water, with concentrations between 10\% and 40\%, can be modeled by the Bingham plastic model~\citep{zengeni-bingham}. Since the concentrations of particles in the present study fall in this range, the Bingham plastic model is used to model the rheological behavior. The constitutive equation of the model is shown in Equation \ref{eqn:binghammodel}. The viscosity of the slurry is expressed as a sum of a constant value ($\mu_{b}$) and a term regarding the yield stress ($\tau_{y}$), which indicates the initial amount of stress required to make the fluid flow. 
\begin{center}
	\begin{equation}  \tag{4}\label{eqn:binghammodel}
		\mu_{l}=\mu_{b}+\frac{\tau_{yield}}{\dot{\gamma}}
	\end{equation}  
\end{center}

\subsection{Computational domain and boundary conditions}

\begin{figure*}[!ht] 
	\centering
	\includegraphics[width=\columnwidth]{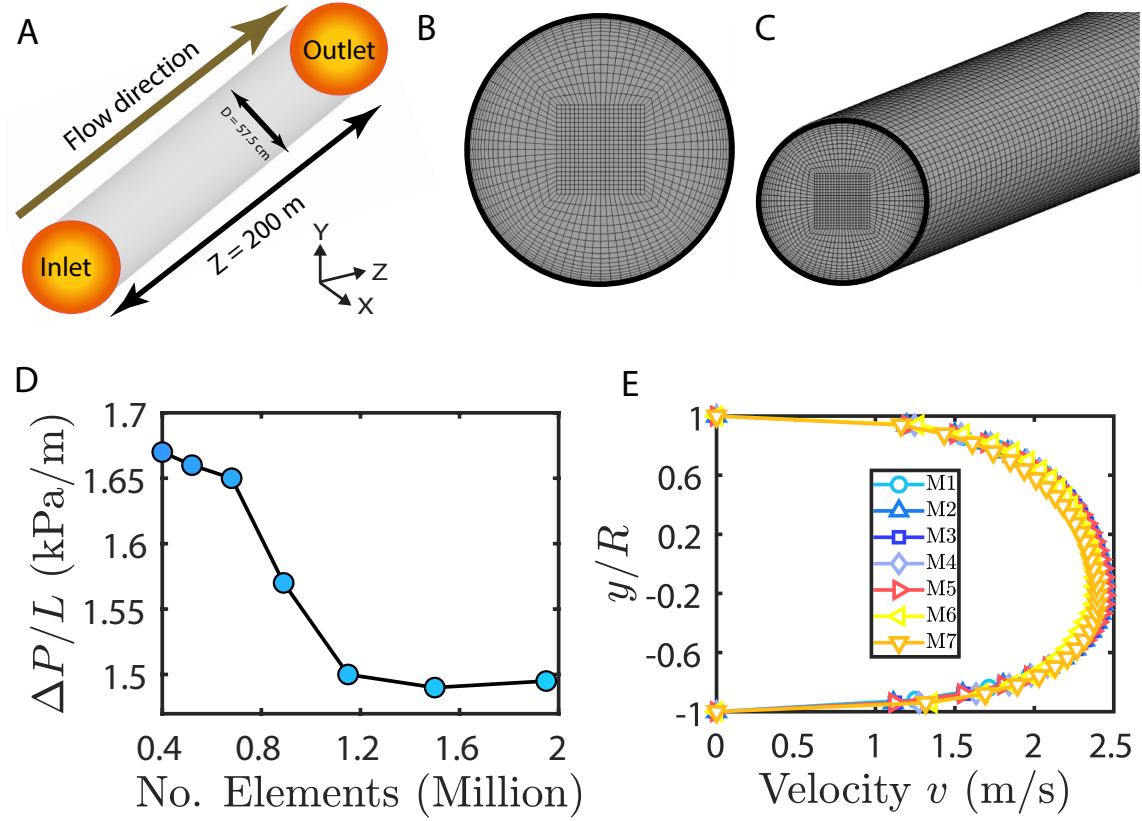} 
	\renewcommand{\captionfont}{\footnotesize}
	\caption{ (A) Simulation domain, (B) cross-sectional view of mesh structures, (C) mesh structure along the length of pipeline, (D) pressure gradient variation for different set of meshes ($M_{1}$ to $M_{7}$), and (E) velocity profile for different set of meshes ($M_{1}$ to $M_{7}$)  at Z = 175 $m$.}
	\label{f01}
\end{figure*}

The computational domain is set as a cylindrical pipeline with a diameter (D) of 0.575 $m$ and a length (Z) of 200 $m$ as illustrated in Figure \ref{f01} A.

The inlet boundary condition is set with a uniform velocity profile for slurry transport process and the velocity is the same for all phases while the representative inlet volume fraction of secondary phases are defined using the PSD with the coal content equal 50\% weight fraction and the sand having 3\% weight fraction, meanwhile, the pressure at the outlet is set to the atmospheric pressure. A wall boundary with a no-slip condition indicates a velocity of zero over the wall surface. The bed formation is achieved by setting up both the inlet velocities of water and solids to zero, and imposing a zero mass flow condition at the outlet. This CFD model configuration leads to the sedimentation of the suspended solids to settle under gravity. After approximately 6 s, a sediment bed formation is observed in the pipeline. After the sand and coal particles bed is formed, the flushing process initiates which has the same boundary conditions though the total inlet weight fraction of the coal and sand is set to zero and pure water flows through the pipe with sand and coal particles distributed in pipeline, the fluid and domain physical properties are given in Table \ref{tab:properties}. In addition to achieving the fully developed flow inside the pipe during transportation, the length of the domain is more than 90D in all simulations \citep{fullydevelopement}.\\

\begin{table}[!ht]
	\renewcommand\thetable{7}
	
	\centering
	\caption{Material properties used in CFD simulations.}
	\label{tab:properties}
	\renewcommand{\arraystretch}{1}
	\begin{centering}
		\renewcommand{\arraystretch}{1}  
		\setlength{\tabcolsep}{4pt}  
		\begin{tabular}{l l }
			
			\hline
			Parameter & Value \\
			\hline
			Slurry transport & \\
			\hline
			Pipe diameter(m) & 0.575  \\
			Pipe length (m) & 200\\
			Sand phases diameter ($\mu$m) & 547, 218\\
			Sand phases volume fractions & 0.013, 0.011\\
			Density of the solid particles ($kg/m^3$)  &  2300\\
			Coal phases diameter ($\mu$m) & 161, 290, 411, 672\\
			Coal phases volume fractions & 0.023, 0.11, 0.21, 0.075\\
			Density of the coal particles ($kg/m^3$)  &  1420\\
			Carrier density ($kg/m^3$) & 1071.3\\
			Bingham viscosity $\mu_b$ ($pa$.s) & 0.0526 \\
			Bingham yield stress $\tau_{yield}$ ($pa$) & 2.755 \\
			\hline
			Flushing & \\
			\hline
			Water density ($kg/m^3$) & 998.19\\
			Water viscosity ($pa$.s) & 0.001139 \\
			Sand and coal inlet volume fraction & 0 \\
			\hline
			
		\end{tabular}
	\end{centering}
\end{table}

\subsection{Modeling of the flushing process}

The finite volume method (FVM) based ANSYS Fluent solver 2022 R2 is used to perform the unsteady state simulations. Table \ref{tab:modelss} provides the solver settings and schemes used in the mixture model.  The transient simulations involve the discretization of governing equations in both space and time. Transient discretization also works on spatial discretization, similar to steady-state simulations using the second-order implicit scheme for stability~\citep{Ansys}. The turbulence intensity and hydrodynamic diameter for all phases are defined as 10\% and 0.575 $m$, at the inlet for both sanding and flushing, respectively.  The volume fraction is computed using the quick method, and specified relaxation factors are applied for pressure, momentum, and volume fraction. 

A specific range of physical flow time is always required to obtain realistic flow conditions. In the current study, the sanding and flushing processes have been run for 200~$s$ and 300~$s$, respectively. All the simulations are run by high performance computing system, Compute Canada, using the Graham cluster with 44 CPUs for 10 days. The 0.001 s times step size is adopted to perform the simulation for slurry transport with 200000 time steps, and flushing process with 300000 time steps. Moreover, the convergence criteria are fixed to 10$^{-5}$ for the resolution of continuity and momentum equations. The post-processing of the results is performed by creating several cross-sections over the longitudinal distance of Z = 60 $m$, Z = 75 $m$, Z = 125 $m$, Z = 140 $m$, and Z = 175 $m$. The plane at Z = 0 $m$ and Z = 200 $m$ are the inlet and outlet, respectively. The distribution of different phases and the center line variation over the cross-sectional plane are examined.\\

\begin{table}[!ht]
	\renewcommand\thetable{8}
	
	\centering
	\caption{Solver conditions used in CFD simulations.}
	\label{tab:modelss}
	\renewcommand{\arraystretch}{1}
	\begin{centering}
		\renewcommand{\arraystretch}{1}  
		\setlength{\tabcolsep}{4pt}  
		\begin{tabular}{l l }
			
			\hline
			Model & Scheme \\
			\hline
			Multiphase model  & Mixture  \\
			Turbulence model & Reynolds stress model\\
			Turbulence dispersion & \citet{burns2004favre}\\
			Turbulence multiphase  &  Mixture\\
			Carrier-solid drag model  & Gidaspow\\
			Carrier-coal drag model  & Gidaspow\\
			Carrier-bubble drag model & Symmetric~\cite{Sekhar2023POF,yiyi} \\
			Scheme & SIMPLE\\
			Gradient & Least Squares Cell Based\\
			Pressure  & PRESTO \\
			Momentum  & Second order upwind\\
			Volume fraction & Quick \\
			Turbulent kinetic energy & Second order upwind \\
			Turbulent dissipation rate & Second order upwind \\
			Transient formation & Second order implicit \\
			Carrier fluid shear condition  & No-slip\\
			Carrier viscosity (slurry transport) & Bingham plastic model\\
			Carrier viscosity (flushing) & Newtonian model\\
			Fraction packing limit & 0.63~\cite{Ansys}\\
			Angle of internal friction & 30~\cite{Ansys}\\
			Particle-particle COR (e) & 0.90~\cite{ppcollision}\\ 
			Time step & 0.001 s \\
			Number of time steps (slurry transport) & 200,000\\
			Number of time steps (flushing) & 300,000\\
			
			\hline
			
		\end{tabular}
	\end{centering}
\end{table}

\subsection{Computational mesh selection}

The effect of mesh density on flow characteristics is explored systematically using the seven different sets of mesh. The number of nodes for each mesh are as follows: 0.40 million, 0.52 million, 0.68 million, 0.89 million, 1.15 million, 1.50 million, and 1.95 million. The mesh type in all the meshes is hexahedron. The mesh quality was evaluated using minimum orthogonal quality (0.72), maximum skewness (0.7), and aspect ratio range (4.4 - 19.7) to select the optimal mesh. The domain also considers 33 boundary layers with inflation near the wall regions to ensure an acceptable $\begin{alignedat}{8} y^{+} \end{alignedat}$ (34.2), while capturing the near-wall effect. Figures ~\ref{f01} B-C show the 3D computational structured M6 mesh over the cross-section and along the length of the pipe employed in the study. It is important to note that the simulation results revealed that the mesh is refined near the wall regions, where most solids at the pipeline invert are accumulated. The mesh refinement could be slightly helpful, though the mesh study indicated that the differences and error would be less than 2\%.

The performance of the computational domain is checked by comparing pressure gradient in the pipeline and velocity profiles along the vertical reference line at Z = 175 $m$. Figures ~\ref{f01} D-E present the pressure gradient and velocity on the mid vertical line for all sets of mesh, respectively. It can be seen from Figure ~\ref{f01} D that the pressure gradient change for the mesh M1 to M7 shows a constant trend after Mesh M6, which consists of 1.50 million elements. The results presented in Figure ~\ref{f01} E demonstrate that the velocity profile is nearly identical between the 1.15 million, 1.50 million, and 1.95 million element meshes. Consequently, mesh M6 with 1.50 million elements is used for further investigation in the study.\\

\begin{figure*}[!ht] 
	\centering
	\includegraphics[width=0.80\columnwidth]{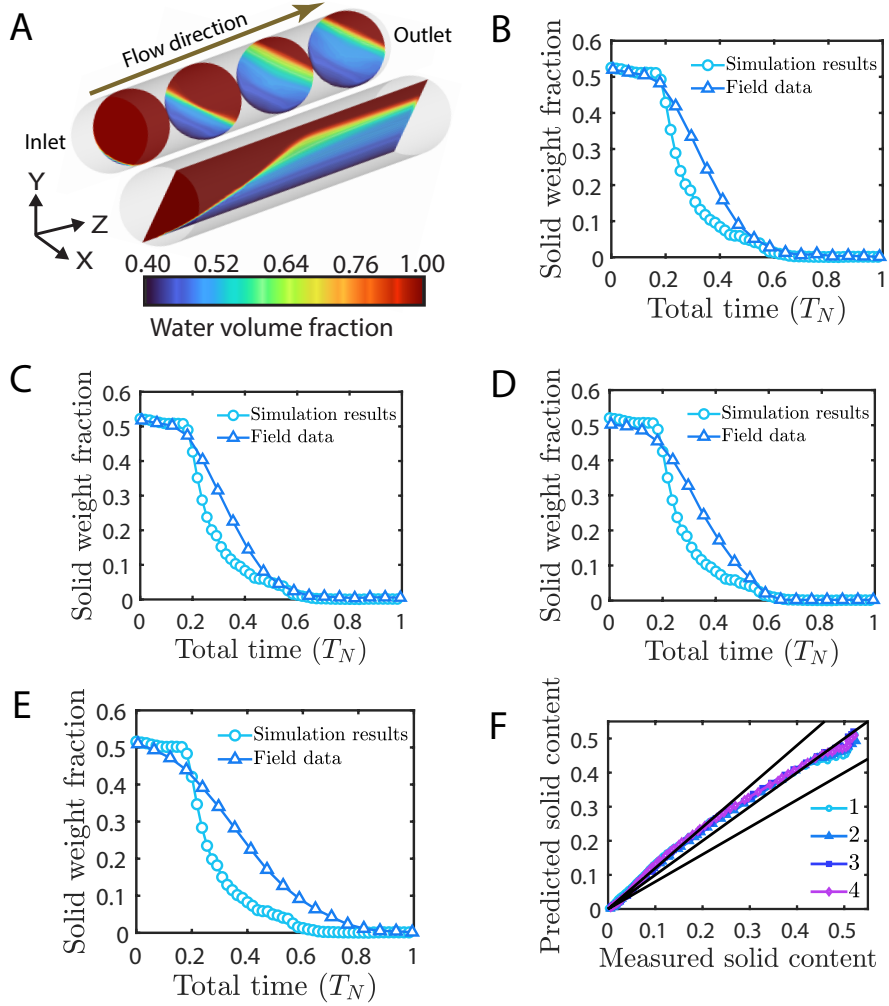} 
	\renewcommand{\captionfont}{\footnotesize}
	\caption{ (A) The carrier fluid volume fraction contour illustration of pipeline interface at Z = 25 $m$, Z = 75 $m$, Z = 125 $m$ and Z = 175 $m$ distance from the inlet, as well as the longitudes at Y-Z (x=0) plane representation of the carrier fluid volume fraction in the pipeline at t = 60 $s$ (t/T = 0.2) of Case 1. This reveals the carrier fluid dynamics during flushing. The concentration of solid content at the outlet interface. Flushing process Validation with different data sets (B) pressure station 1, the Case 1 (C) pressure station 2, the Case 2 (D) pressure station 3, the Case 3 (E) pressure station 4, the Case 4 and (F) parity plot of each Case which would indicate over 90\% of the simulation data has less than 20\% error while the average error is $\pm$ 15\%.}
	\label{fig:6}
\end{figure*}

%

\section{Characterization of flushing efficiency}

The predicted solid weight fraction is defined as the ratio of the difference between the mixture and carrier fluid flux, to the mixture flux at the outlet section. It can be calculated as follows:
\begin{center}
	\begin{equation}  \tag{5}\label{eqn:outletconcentration}
		C_{s}=\frac{\int\int\rho_{mixture}\Vec{v}_{m}.d\Vec{A}-\int\int\rho_{carrier-fluid}\Vec{v}_{cf}.d\Vec{A}}{\int\int\rho_{mixture}\Vec{v}_{m}.d\Vec{A}}|_{outlet}
	\end{equation}  
\end{center}

The normalized time is defined as the ratio of the time to the total time. It can be calculated as follows: 

\begin{center}
	\begin{equation}  \tag{6}\label{eqn:normalizedtime}
		T_{N}=t/T
	\end{equation}  
\end{center}


To investigate the flushing process, several key parameters like pipeline cleanness (CL), cleanness efficiency (CE), deposit occupation (DO), normalized deposit occupation (NDO), and carrier-fluid consumption (CFC) are also examined. In the system, all secondary solid phases are considered as deposit and dirtiness, while the carrier fluid is used to flush the solid from the pipeline is corresponds to cleanness. Thus, the average carrier fluid volume fraction in the pipeline is interpreted as the CL of the system. Equation \ref{eqn:cleanness} is used to calculate the CL. The CL is estimated as the integration of carrier fluid volume fraction over the volume of the pipeline, and divided by the pipeline volume. 

\begin{equation}  
	\tag{7}\label{eqn:cleanness}
	\text{CL(t)~=~} \overline{\phi}_{carrier}= \frac{1}{V} \int\int\int_{c.v} \phi_{carrier}(t) dV
\end{equation}  

To evaluate the effectiveness of the flushing process, it is essential to observe the fraction of solids removed from the pipeline over time. The CE is defined as the fraction of the solid content removed from the pipeline to the initial solid content in the pipeline over the designated period. It can be calculated using the equation below:\\

\begin{equation}  \tag{8}\label{eqn:CE}
	CE = \frac{CL(t)-CL(t_{0})}{1-CL(t_{0})}
\end{equation}  

The DO is a parameter that measures the specific secondary solid phases throughout the pipeline. DO is defined as the average volume fraction of the secondary solid phase present in the pipeline and it can be calculated as follows:
\begin{equation}  
	\tag{9}\label{eqn:contaminent}
	DO_{i}(t) = \overline{\phi}_{Deposit_{i}} = \frac{1}{V} \int\int\int_{c.v} \phi_{Deposit_{i}}(t) dV
\end{equation}

Further, to compare the deposit removal, DO is normalized and reported in the form of NDO. NDO is defined as the DO divided by the deposit present at the start of flushing. It can be calculated as the following equation:

\begin{equation}  \tag{10}\label{eqn:contaminentnorm}
	NDO_{i}(t) = \frac{DO_{i}(t)}{DO_{i}(t_{0})}
\end{equation}  

Another important parameter during the flushing process could be the consumption of the carrier fluid, that is, flushing water. To calculate the $CFC^{c}_{n}$, it is essential to identify the desired cleanness of the slurry transportation pipeline.  $CFC^{c}_{n}$ is used to represent the amount of water required to clean the pipeline, where $c$ and $n$ indicate the cleanness and degree of cleanness, respectively. The $CFC^{c}_{n}$ can be estimated by using the equation below:
\begin{center}
	\begin{equation}  \tag{11}\label{eqn:cfc}
		CFC^{c}_{n}= \int_{t_{0}}^{t_{n}^{c}} \Vec{A}.\Vec{v}_{cf}(t)dt 
	\end{equation}  
\end{center}

$t_{n}^{c}$ is the time required to reach CL (cleanness time) of $n$, which can be calculated using the
equation as follows:

\begin{equation}  \tag{12}\label{eqn:tnc}
	n = \overline{\phi}_{carrier}(t_{n}^{c})= \frac{1}{V} \int\int\int_{c.v} \phi_{carrier}(t_{n}^{c}) dV
\end{equation}

Moreover, the normalized carrier fluid consumption ($NCFC^{c}_{n}$) is defined as the $CFC^{c}_{n}$ divided by the pipeline volume and can be represented as follows:

\begin{equation}  \tag{13}\label{eqn:ncfc}
	NCFC^{c}_{n}=  \frac{CFC^{c}_{n} }{V_{pipeline}}
\end{equation}

\section{Results and discussion}

\subsection{Model validation}

Figure \ref{fig:6} A shows an example of the CFD model proposed by this paper. The present model is able to capture the 3D information of the water volume fraction and shape of the solid bed in an enclosed pipeline.  To verify the efficacy of the developed numerical model,  the numerical model is validated by comparing the predicted and in situ measured solid weight fraction.

Figures \ref{fig:6} B-E presented the comparison of the predicted solid weight fraction from CFD simulation presented in Table \ref{tab:validationCases} at the outlet section with the in-situ measurement over the pipeline outlet section. The instantaneous flushing time is normalized with the total time of process and used to compare the rate of flushing.

It can be seen from Figures \ref{fig:6} B-E that at different operating conditions the solid content is found to decrease with the time. The decreasing trend of predicted solid content over time is found to agree with the in-situ measurements. Further, to emphasize the validity of the developed numerical model a parity plots is developed and reported in Figure \ref{fig:6} F. The predicted solid weight fraction is found to vary by an average error of about $\pm$15\% compared to the in-situ field data. The reason for the difference between the simulation results and field data is that the solid-solid interactions are neglected in the mixture model. This indicates the solids do not have an impact on one another and this will result in the difference of the field data and simulation results with an average error of $\pm$15\% , which is an accepted range for such a large-scale complex problem. The reasonable agreement with the field data confirms that the developed model is capable of predicting the flow parameters for the industrial flushing process conditions. The CFD model predictions are further extended for the study of the parametric influence on the slurry flushing process.

In the following sections, the effects of water velocity, bubble size, and bubble fractions on the mechanisms and performance of the flushing process are analyzed.

\subsection{The solid bed formation during slurry transport}

\begin{figure}[!ht] 
	\centering
	\includegraphics[width=0.80\columnwidth]{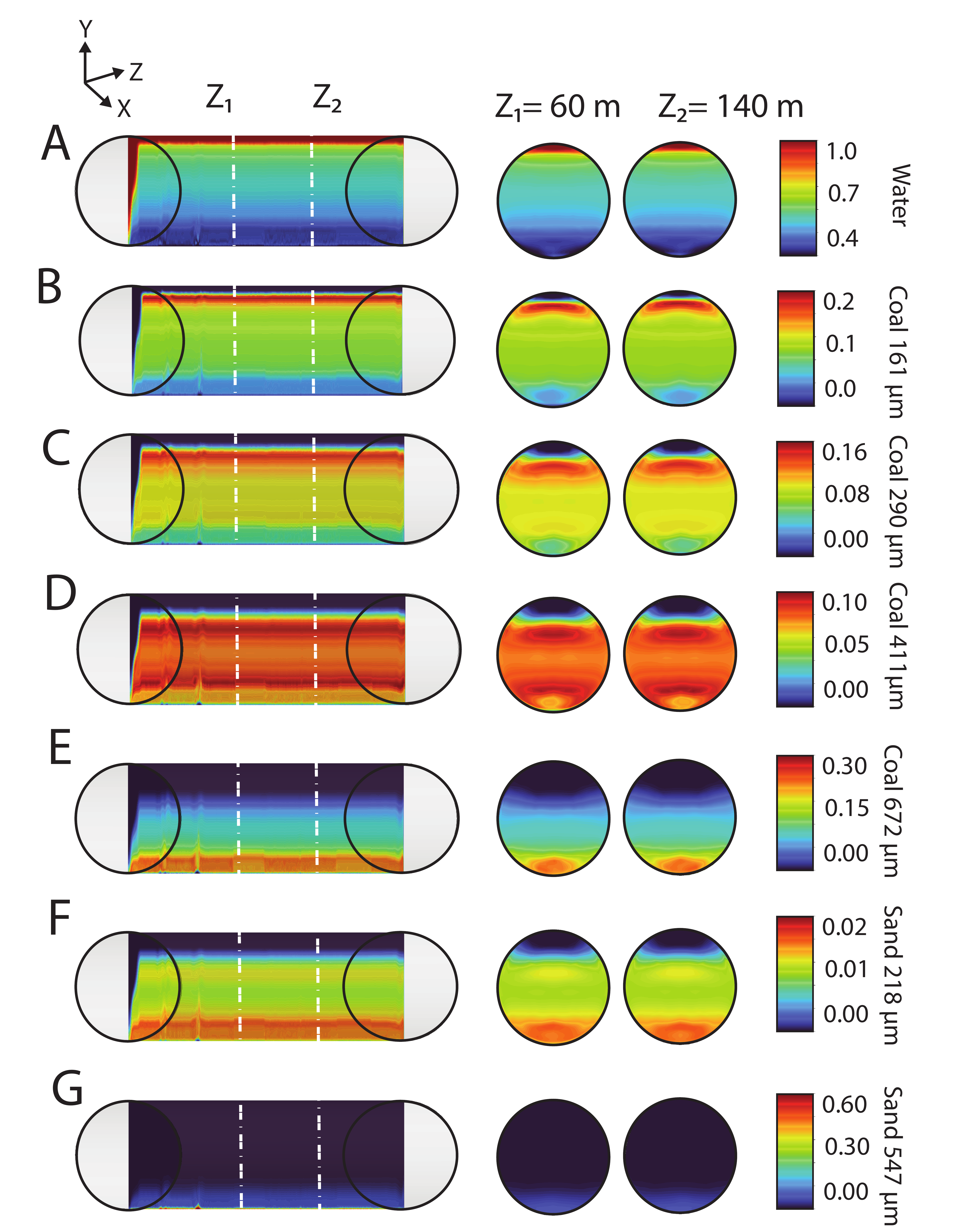} 
	\renewcommand{\captionfont}{\footnotesize}
	\caption{The contour plots of volume fraction of (A) the water, (B) the 161 $\mu m$ coal, (C)  the 290 $\mu m$ coal (D)  the 411 $\mu m$ coal
		(E)  the 672 $\mu m$ (F), the 218 $\mu m$ solid and (G)  the 547 $\mu m$ solid in the longitudinal section and two cross sections (Z1 = $60~m$, Z2 = $140~m$).}
	\label{fig:solidbed}
\end{figure}

The flushing process begins after the slurry transportation. The slurry transportation serves as the flushing initial condition, which continues until the system reaches a pseudo-steady state where flow and sediment dynamics stabilize. Figure \ref{fig:solidbed} shows the distribution of the water fraction (Figure  \ref{fig:solidbed} A) and the solid content distribution  (Figure \ref{fig:solidbed} B-G)in both longitude section and two cross sections when the slurry transport is at a pseudo steady state condition in the pipeline.

Figure \ref{fig:solidbed} A shows that the carrier is dilute at the top during the slurry transport. In contrast, the slurry becomes more concentrated at the pipeline invert due to the solid content inertia forces. Figures \ref{fig:solidbed} B-G reveal the coal and sand particle distribution along two intersections of the pipeline ($Z_1$ and $Z_2$) and in longitude. The smaller particles mostly occupy the top section, while their larger counterparts, due to their greater inertia force, are concentrated deeper in the stratified slurry flow~\cite{peker2011solid}.

\subsection{Effect of water velocity }
\begin{figure}[!ht] 
	\centering
	\includegraphics[width=0.85\columnwidth]{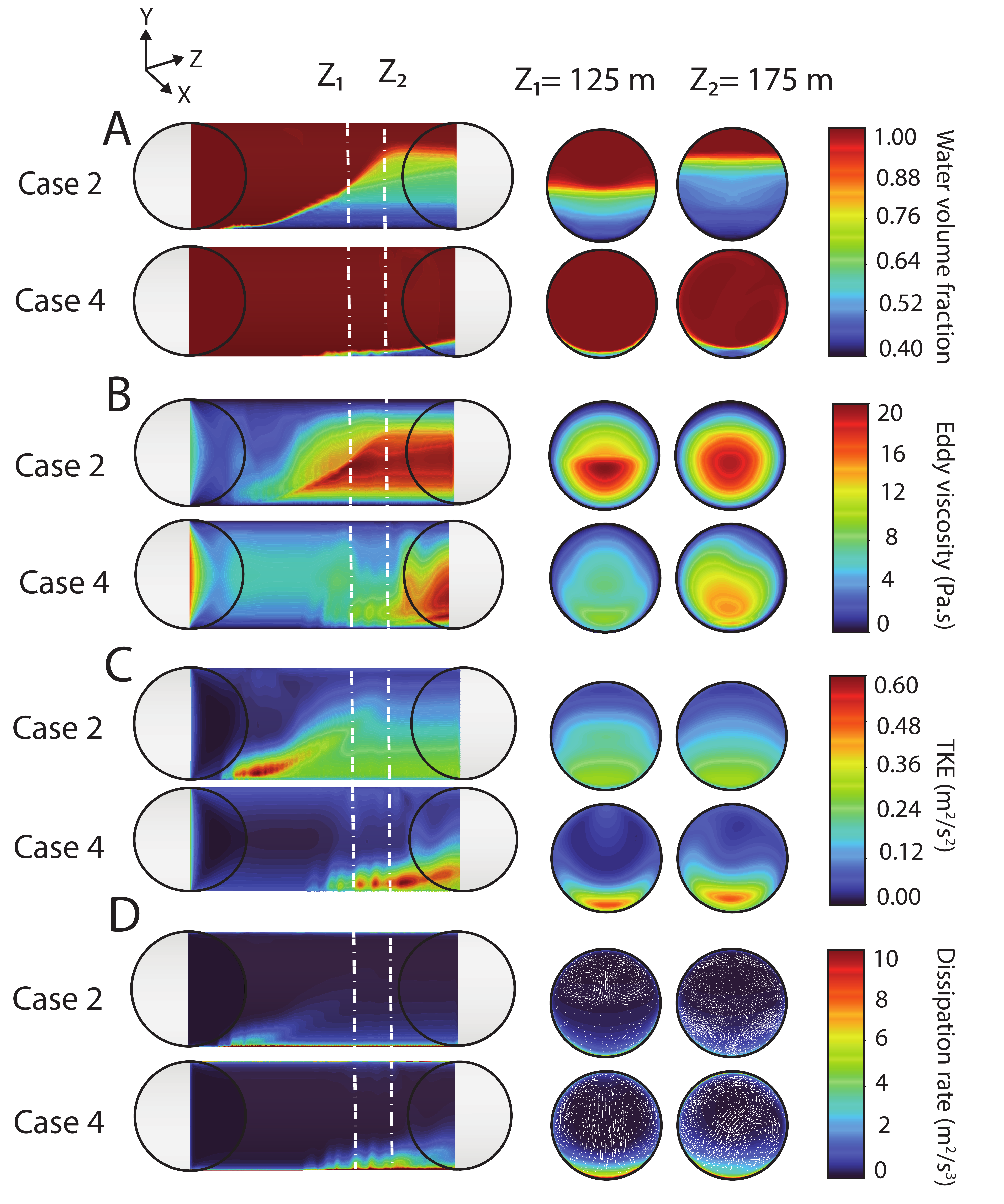}
	\renewcommand{\captionfont}{\footnotesize}
	\caption{The contour plots of (A) the water volume fraction, (B) eddy viscosity, (C) turbulent kinetic energy, and (D) dissipation rate in the longitudinal section and two cross sections ($Z_1=125~m$, $Z_2=175~m$) for case 2 ($v_{cf}=2.88~m/s$), and case 4 ($v_{cf}=4.88~m/s$) at $t=60~s$ $ (t/T=0.2)$.}
	\label{fig:30}
\end{figure}

Figure \ref{fig:30} A-D shows the water volume fraction and characteristics of turbulence in both longitudinal section and two cross sections at $t= 60~s$ for two water velocities $v_{cf}$ at $2.88~m/s$ and $4.88~m/s$. The water volume fraction in the pipeline is larger at higher velocity, which indicates that higher water velocity facilitates the removal of sediments and shortens the required flushing time. \\

\begin{figure*}[!htbp] 
	\centering
	\includegraphics[width=\columnwidth]{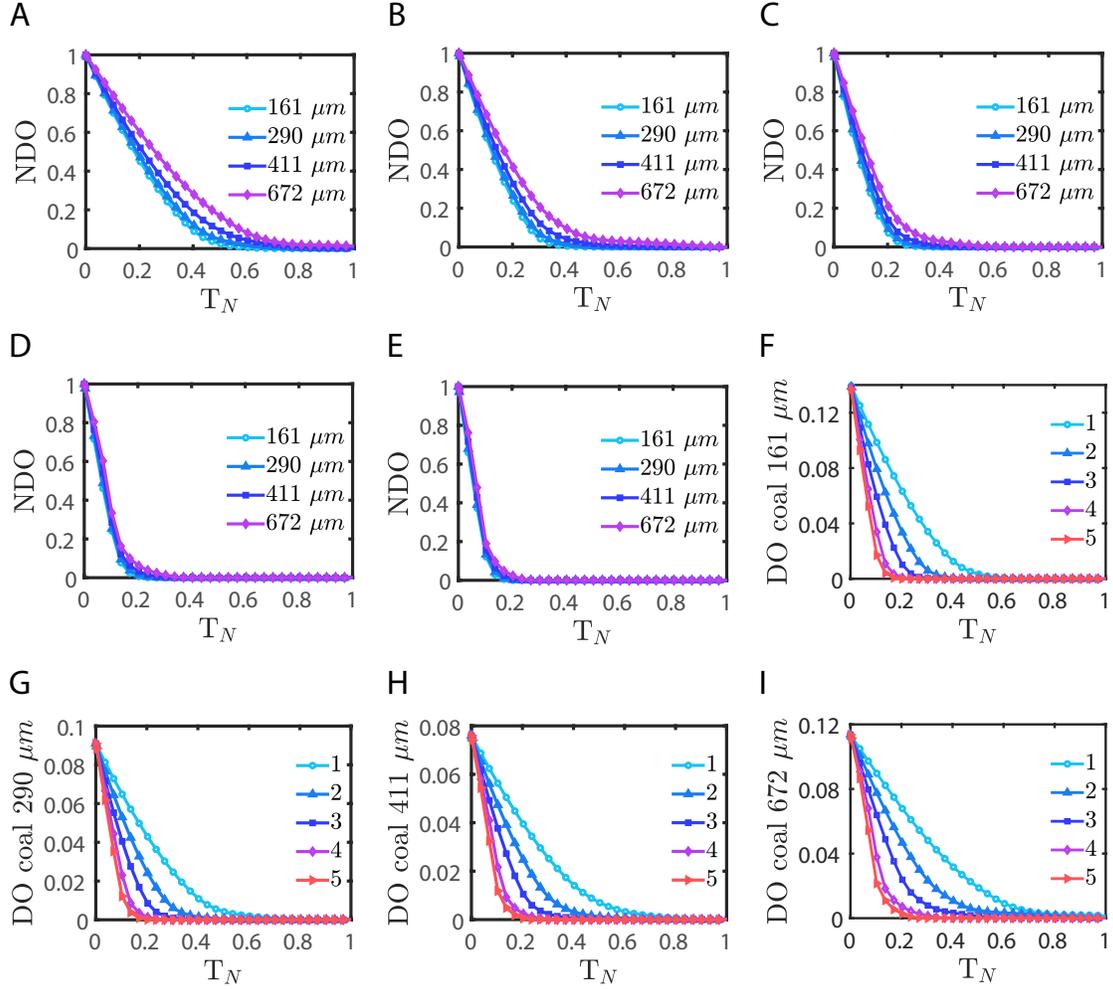} 
	\renewcommand{\captionfont}{\footnotesize}
	\caption{The NDOs of different sized particles under five water velocity (A) case 1 ($v_{cf}=1.88~m/s$), (B) case 2 ($v_{cf}=2.88~m/s$), (C) case 3 ($v_{cf}=3.88~m/s$), (D) case 4 ($v_{cf}=4.88~m/s$), (E) case 5 ($v_{cf}=5.88~m/s$), and the DO factors at five different water velocities (case 1,2,3,4,5) for coal particles with size (F) $161 ~\mu m$, (G) $290 ~\mu m$, (H) $411 ~\mu m$, and (I) $672 ~\mu m$. }
	\label{fig:f08}
\end{figure*} 

In addition, figure \ref{fig:30} A shows the 3D shape of the solid bed, from which the pattern that solid particles are suspended and removed can be inferred. For case 2 ($v_{cf} = 2.88~m/s$), in the longitudinal section, a ramp is formed at the windward side in the solid bed. This is due to the spatial distribution of particles of different sizes in the solid bed, i.e., the size of particles is smaller as it gets closer to the surface of the solid bed. Since the smaller particles at the surface has smaller inertia, they are easier to move with the water flow, piling up at the windward side and forming a bump at the top the ramp. The larger particles closer to the bottom of the solid bed are then exposed to water flow, resulting in a stratified pattern in the particle removal process. In the cross sections at the ramp ($Z_1$), the thickness of the solid bed is the smallest at the central axis and gets larger close to the wall of the pipe. This is because the flow is faster at the central axis than close to the wall due to the shear forces between the solid bed and wall of the pipe. These observations can also be seen for case 4~($v_{cf} = 4.88~m/s$), although the ramp is not as obvious since most of the particles are removed at this time.

\begin{figure*}[!htbp] 
	\centering
	\includegraphics[width=\columnwidth]{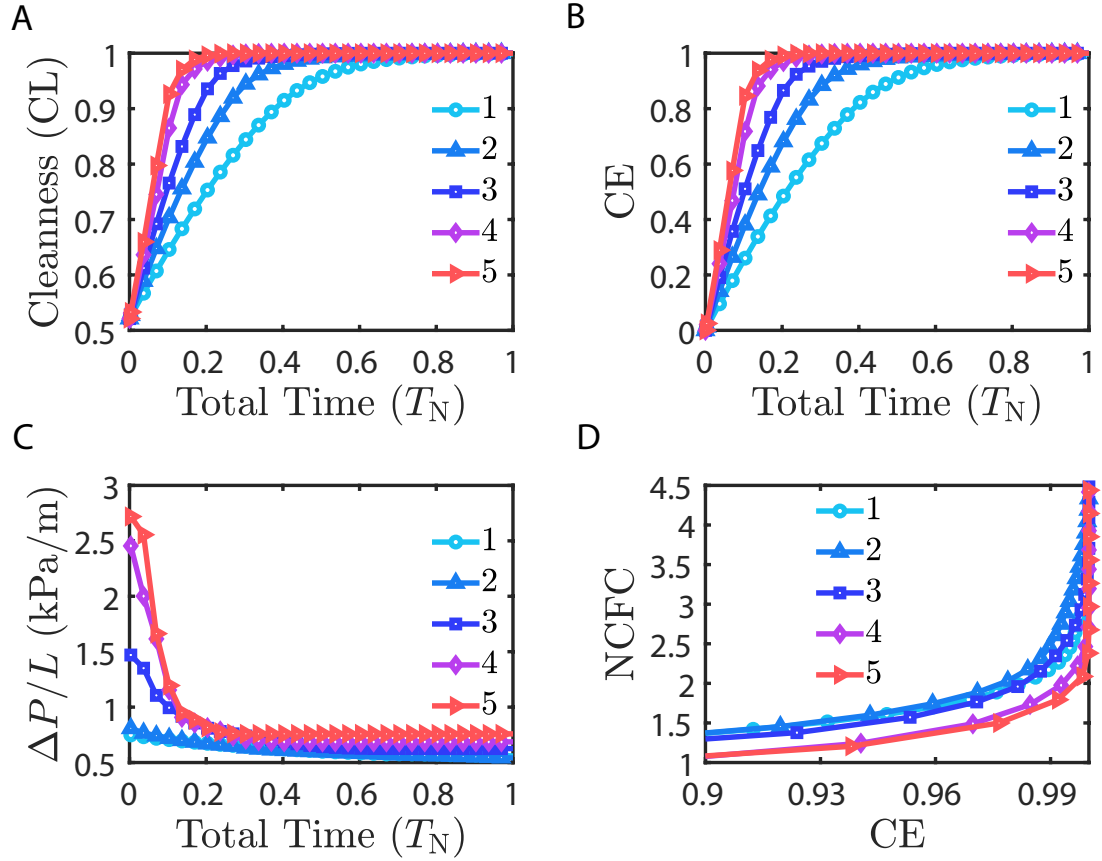}  
	\renewcommand{\captionfont}{\footnotesize}
	\caption{ (A) cleanness, (B) cleanness efficiency, (C) pressure gradient along pipeline, and (D) normalized water consumption with respect to cleanness for case 1 ($V_{cf}=1.88 ~m/s$), 2 ($V_{cf}=2.88 ~m/s$), 3 ($V_{cf}=3.88 ~m/s$), 4 ($V_{cf}=4.88 ~m/s$), and 5 ($V_{cf}=5.88 ~m/s$). }
	\label{fig:f09}
\end{figure*}

The mechanism of the stratified removal patterns can be validated by the turbulent characteristics shown in figure \ref{fig:30} B-D. Figure \ref{fig:30} B shows the eddy viscosity in the pipeline. The eddy viscosity is related to the drag forces between different phases, which reflects the strength of the interaction between water and solid phases in the model. The higher the eddy viscosity, the stronger the drag forces are, indicating that more particles are suspended by the water flow. For case 2, the eddy viscosity is the highest at the center of the solid bed, which indicates strong mixing and formation of vortices at this region. The vortices help the water flow to penetrates the solid bed and suspend the particles. This is in accordance with figure \ref{fig:30} A where the water volume fraction is moderately high at the center of the solid bed. The eddy viscosity gets smaller closer to the surface of the solid bed because the concentration of particles gets smaller and it is easier for water to flow through smaller particles. The eddy viscosity drops almost to zero at the bottom of the solid bed closer to the pipe of the wall, which indicates poor mixing in this region. This is because the sizes of the particles are large at the bottom of the solid bed, which makes it hard for water to flow through. As a result, the largest particles at the bottom of the solid bed is the hardest to remove. Similar pattern is also observed for case 4, which is at a much later stage of flushing~\cite{wang2022dynamic}.

Figure \ref{fig:30} C and D shows the turbulent kinetic energy and dissipation rate, respectively. The turbulent kinetic energy is the highest at the windward side closer to the bottom of the ramp where large size particles lies, indicating larger velocity fluctuations in this region~\cite{gai2020particles}. This may be attributed to the the phenomenon known as particle creeping, where flow attempts to suspend large particles but it fails. Meanwhile, due to particle creeping, the turbulent dissipation rate is also the highest in the this region. These observations are in accordance with the low eddy viscosity in this region in figure \ref{fig:30} B, showing limited effect of flow on suspending large particles~\cite{wang2019large}.

In the cross-section plot of figure \ref{fig:30} D, the vortices are shown by the white arrows. While at the early stage of flushing the vortices are symmetrical with respect to the central axis (case 2), they become oblique at the late stage (case 4). This indicates that the turbulence in flushing process is in essence anisotropic and it becomes more apparent when most of the particles are removed from the pipe. In addition, the asymmetry of the vortices may also be attributed to the way the solid bed is generated initially. Since the particles of different sizes are deposited in the pipeline during slurry transport, the spatial distribution of particles is not symmetrical, which better reflects the real-world scenario. \\

Figure \ref{fig:f08} A-E shows the effect of water velocity on the efficiency of removing particles of different sizes. At low water velocities (Figure \ref{fig:f08} A and B), smaller particles are removed faster from the pipeline than their larger counterparts. This indicates that the turbulence is not strong enough to suspend large particles and the large particles may creep. Furthermore, at very low water velocities (Figure \ref{fig:f08} A), the largest particle ($672 ~\mu m$) cannot be fully removed ($NDO>0$ at $t/T=1$). As water velocities increases, rates of removal for both small and large particles get similar. When the water velocity exceeds the critical velocity at which all the particles can be suspended (Figure \ref{fig:f08} D and E), the rates of removal becomes indistinguishable for different sized particles and the solid bed is removed uniformly from the pipeline~\cite{anzoom2024coarse}.

Figure \ref{fig:f08} F-I compares the rate of removal of different particle sizes at five water velocities. It shows that higher water velocities lead to a faster rate of removal for particles of all sizes. Such an effect is more significant on larger particles than on the smaller counterparts. However, the increment in the rate of removal is diminishing. As the water velocity increases past the critical velocity (cases 4 and 5 ($v_{cf} = 5.88~m/s$)), further increases in water velocity may not be efficient in the flushing process.

Figure \ref{fig:f09} shows the effect of water velocity on the overall flushing performance of removing particles of multiple different sizes from the pipeline. Figure \ref{fig:f09} A and B show that higher water velocity results in achieving both a higher CL and a higher CE in a shorter amount of time. However, higher water velocity requires a higher pressure at the input as shown in Figure \ref{fig:f09} C, as well as a larger mass flow rate. As a result, pursuing a faster flushing process at high water velocity can lead to larger water consumption, which makes the process costly. Figure \ref{fig:f09} D compares NCFCs to reach different CEs at different water velocities. It shows that the water consumption to reach a specific CE first increases (from case 1($v_{cf} = 1.88~m/s$) to case 2) and then decreases (from case 2 to case 5) as water velocity increases. When the water velocity is higher than the critical velocity, the amount of change in water consumption is negligible. However, the pressure required at the initial stage is smaller, which may lower the cost of the flushing process. Moreover, as CE gets closer to 1, the water consumption increases dramatically since it is hard to remove the largest particles left in the pipeline, and the area of interaction between water and solid particles gets smaller in the pipeline. As a result, it should be cautious when targeting an ultra clean pipeline in the flushing process.

\subsection{Effect of bubble size}
\begin{figure}[!ht] 
	\centering
	\includegraphics[width=0.85\columnwidth]{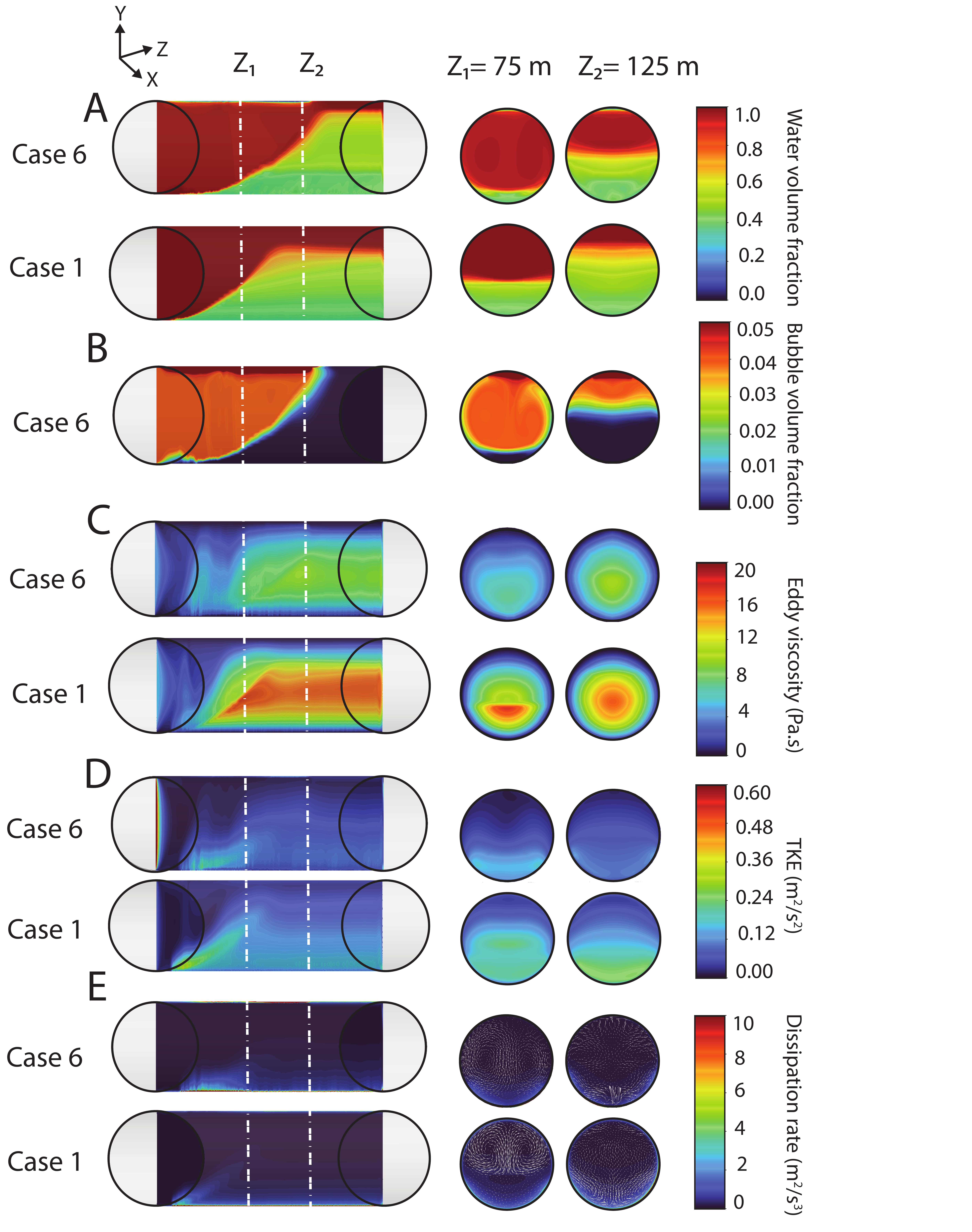}
	\renewcommand{\captionfont}{\footnotesize}
	\caption{The contour plots of (A) the water volume fraction,(B) bubble volume fraction, (C) eddy viscosity, (D) turbulent kinetic energy and (E) dissipation rate in the longitudinal section and two cross sections ($Z_1=75~m$, $Z_2=125~m$) for case 1 (no bubble) and case 6 (bubble volume fraction 0.05 and bubble size 50 $\mu m$) at $t=60~s$ $ (t/T=0.2)$.}
	\label{fig:14}
\end{figure}

The effect of bubble injection on the performance of the flushing process is systematically investigated, as shown in Figure \ref{fig:14} A-E.  

The comparison of water and bubble volume fractions and turbulence characteristics between flushing without bubbles versus flushing with $5\%$ bubbles at $t=60~s$ with the bubble size of 50 $\mu m$ is presented. It is observed that from Figure \ref{fig:14} A and B, the bubbles float at the top of the pipe due to their buoyancy, driven by the density difference between the gas bubbles and concentrated slurry. 

\begin{figure}[!ht] 
	\centering
	\includegraphics[width=0.70\columnwidth]{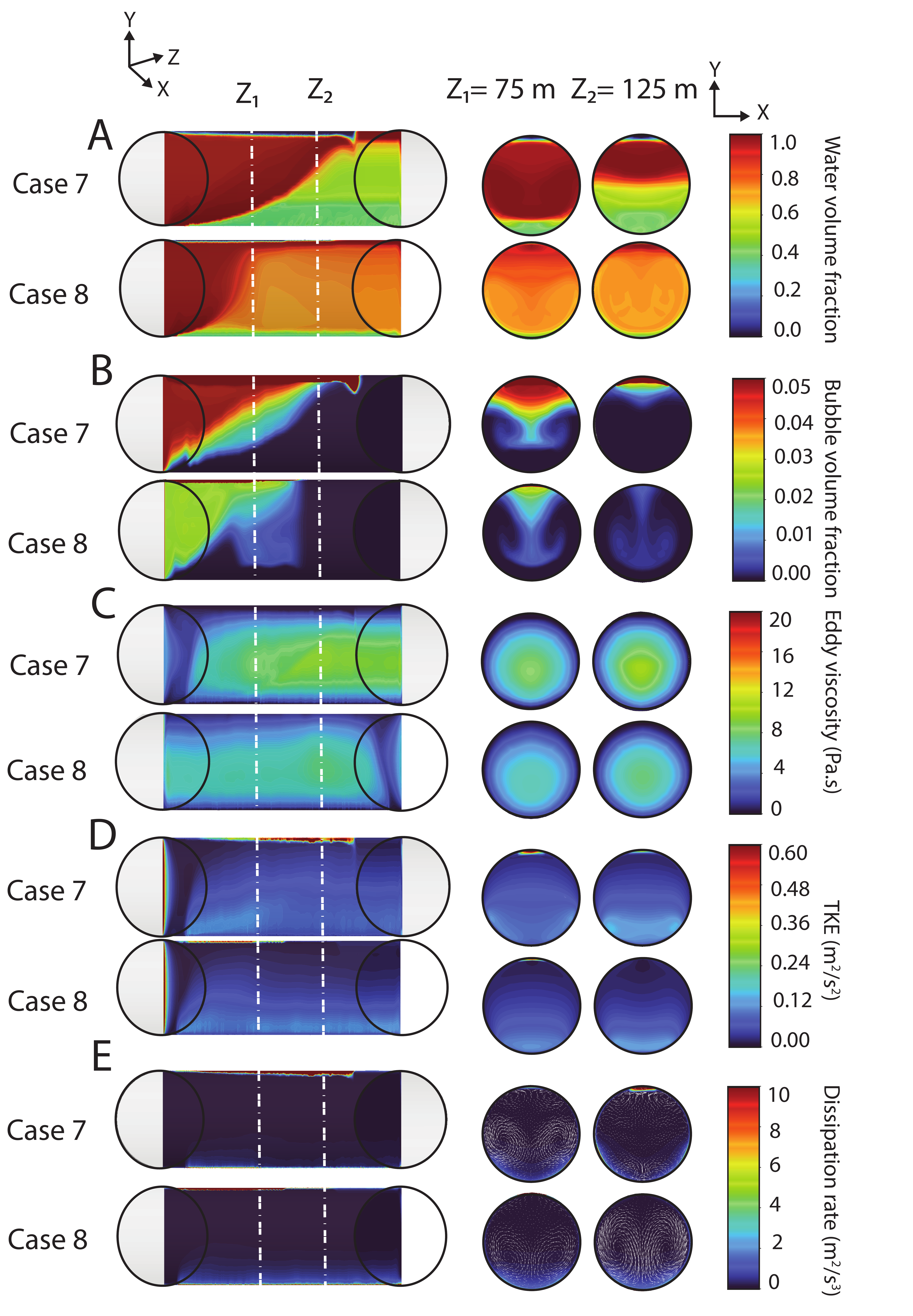} 
	\renewcommand{\captionfont}{\footnotesize}
	\caption{The contour plots of (A) the water volume fraction,(B) bubble volume fraction, (C) eddy viscosity, (D) turbulent kinetic energy and (E) dissipation rate in the longitudinal section and two cross sections ($Z_1=75~m$, $Z_2=125~m$) for case 7 (bubble volume fraction 0.05 and bubble size 150 ~$\mu m$) and case 8 (bubble volume fraction 0.05 and bubble size 1000 ~$\mu m$) at $t=60~s$ $ (t/T=0.2)$.}
	\label{fig:15}
\end{figure}

\begin{figure*}[!htbp]  
	\centering 
	\includegraphics[width=0.80\columnwidth]{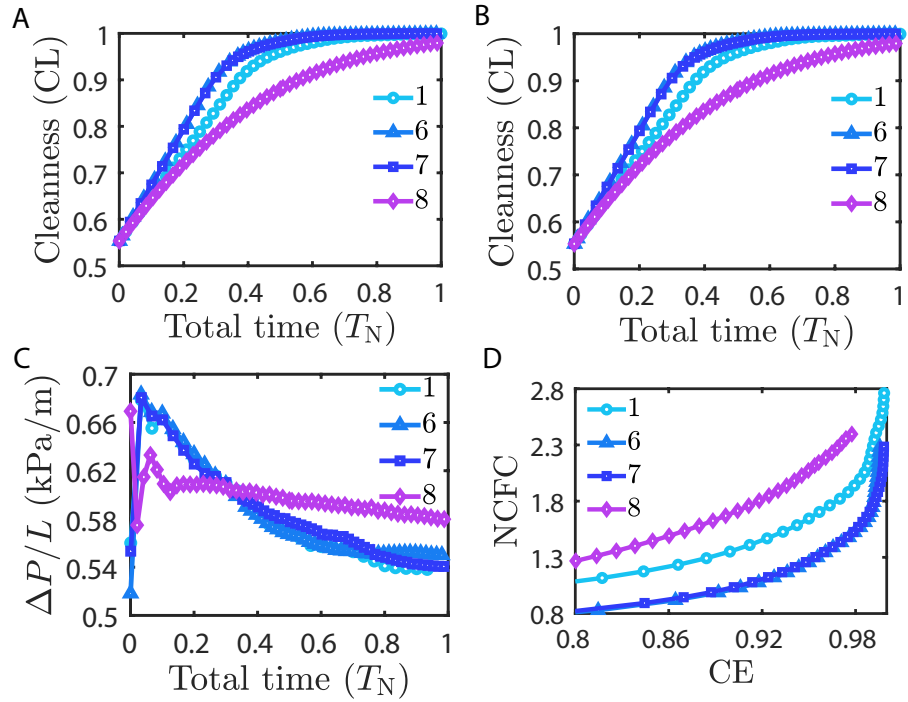}
	\renewcommand{\captionfont}{\footnotesize}
	\caption{ (A) Cleanness, (B) cleanness efficiency, (C) pressure gradient over the pipeline, (D) normalized water consumption with respect to cleanness efficiency for case 1 (no bubble), case 6 (bubble 50 $\mu m$), case 7 (bubble 150 $\mu m$) and case 8 (bubble 1000 $\mu m$).}
	\label{fig:f11}
\end{figure*}
The introduction of gas bubbles into the flushing process significantly enhances the removal of solid content from pipelines, as evidenced by comparative analysis at intersections ($Z_{1} \&~ Z_{2}$). The solid bed removal is accelerated with the presence of gas bubbles, as shown in \ref{fig:14} B, which depicts the interaction between injected gas bubbles and the solid bed. As a result of the bubble imparting a lift force to the solid particles through momentum transfer, the underlying mechanics can be attributed to buoyancy-driven flow in a slurry system~\cite{zhao2023estimation,narayanan1969suspension}.\\


It can be observed from Figure \ref{fig:14} C that the distribution of eddy viscosity is significantly reduced in the presence of bubbles (Case 6). In bubble-free flows (Case 1), intense vortex structures enhance mixing and increase eddy viscosity near the bed. However, bubble injection disrupts these interactions by reducing the liquid phase contact with the solid bed, weakening vortex formation. This suppression of eddies is attributed to the gas-liquid interfacial dynamics, where bubbles introduce coherent upward wakes that locally dampen small-scale turbulence, as shown in Figure \ref{fig:14} D and E. Further, Figure \ref{fig:14} D illustrates the turbulent kinetic energy (TKE), which quantifies the energy associated with turbulent fluctuations. In regions near the solid bed front, bubble injection can contribute to suppresses TKE due to the damping of small-scale eddies. Conversely, at the pipeline crown, where the bubble volume fraction approaches unity, intense bubble-bubble interactions and shear at the gas-liquid interface elevate TKE~\cite{murai2014frictional,gabillet2002experimental}.



In addition, the Figure \ref{fig:14} E demonstrates that the dissipation rate is suppressed by bubbles interacting with solid particles at the solid bed front, which suppresses near-bed eddies and lowers the fluid turbulent viscosity, leading to a reduction of the turbulent dissipation rate. In contrast, at the pipeline crown, accumulated bubbles disrupt eddy circulation, amplify interfacial shear, and generate wake–wake interactions, thereby increasing local dissipation rate~\cite{hoque2024bubble}. Consequently, bubbles enhance particle suspension via momentum transfer and suppress eddy formation, speeding solid removal. Because this suspension-enhancing effect dominates the eddy-damping effect, the overall flushing process is improved.\\

As it can be seen from the aforementioned investigation, bubbles significantly improve the flow process. The impact of bubble size on flushing performance is systematically analyzed in Figure \ref{fig:15} A-E, which presents the distributions of water and bubble volume fractions, as well as turbulent flow characteristics, for bubble diameters ranging from 150~$\mu$m to 1000~$\mu$m. Figure \ref{fig:15} A shows the water volume fraction along the pipeline and across two cross-sections. In Case 8 (1000~$\mu$m bubbles), the solid bed is more dispersed and elevated, forming a thicker bed with slower solid removal rates compared to Case 7 (150~$\mu$m bubbles). This difference is attributed to the greater buoyancy of larger bubbles, which scales with the cube of the bubble diameter ( $F_b$ $\propto$ $d^3$ ). Larger bubbles rise rapidly to the pipeline crown, reducing their interaction with the solid bed and limiting momentum transfer to particles. Figure \ref{fig:15} B highlights the dynamics of large bubbles in Case 8, where their high momentum partially lifts the solid bed, allowing water infiltration and bed dilution. Figure \ref{fig:15} C shows that flows with larger bubbles exhibit lower eddy viscosity, indicating reduced turbulent mixing. This reduction stems from the suppression of small-scale turbulence above the solid bed front, as larger bubbles create coherent wakes that disrupt eddy formation.\\

Figure \ref{fig:15} D shows the distribution of turbulent kinetic energy (TKE). Higher TKE is evident near the pipe crown, where bubble accumulation leads to unsteady, fluctuating motion and interfacial instabilities. In contrast, TKE decreases in the lower regions, particularly upstream of the solid bed, where bubble-particle interactions promote particle lifting and reduce local turbulence intensity. Furthermore, Figure \ref{fig:15} E displays the turbulent dissipation rate. The concentration of bubbles near the top of the pipe increases local dissipation due to enhanced shear and wake interactions, disrupting large-scale eddies. \\

Figure \ref{fig:f11} A-D illustrates the impact of bubble size on pipeline flushing performance for particle removal. Figure \ref{fig:f11} A and B indicates that small bubbles (50 $\mu$m) achieve higher CL and CE faster, while large bubbles (1000 $\mu$m) show initial similarity but decline after ($T_N$ = 0.2) due to solid bed displacement and bubble accumulation at the pipeline crown, reducing effective water flow and solid suspension. Figure \ref{fig:f11} C shows small bubbles maintain a stable pressure gradient that decreases with cleanliness, whereas large bubbles exhibit a rising gradient due to slower solid removal. Further, the CFD results reveals that small bubbles reduce water consumption by 35\%  compared to the no-bubble case, while large bubbles increase it by 23\%, indicating that small bubbles enhance and large bubbles hinder the process as shown in Figure \ref{fig:f11} D~\cite{tao2005role}. The case details are in Table \ref{tab:parameterrange}.

\begin{figure}[!htbp] 
	\centering
	\includegraphics[width=0.70\columnwidth]{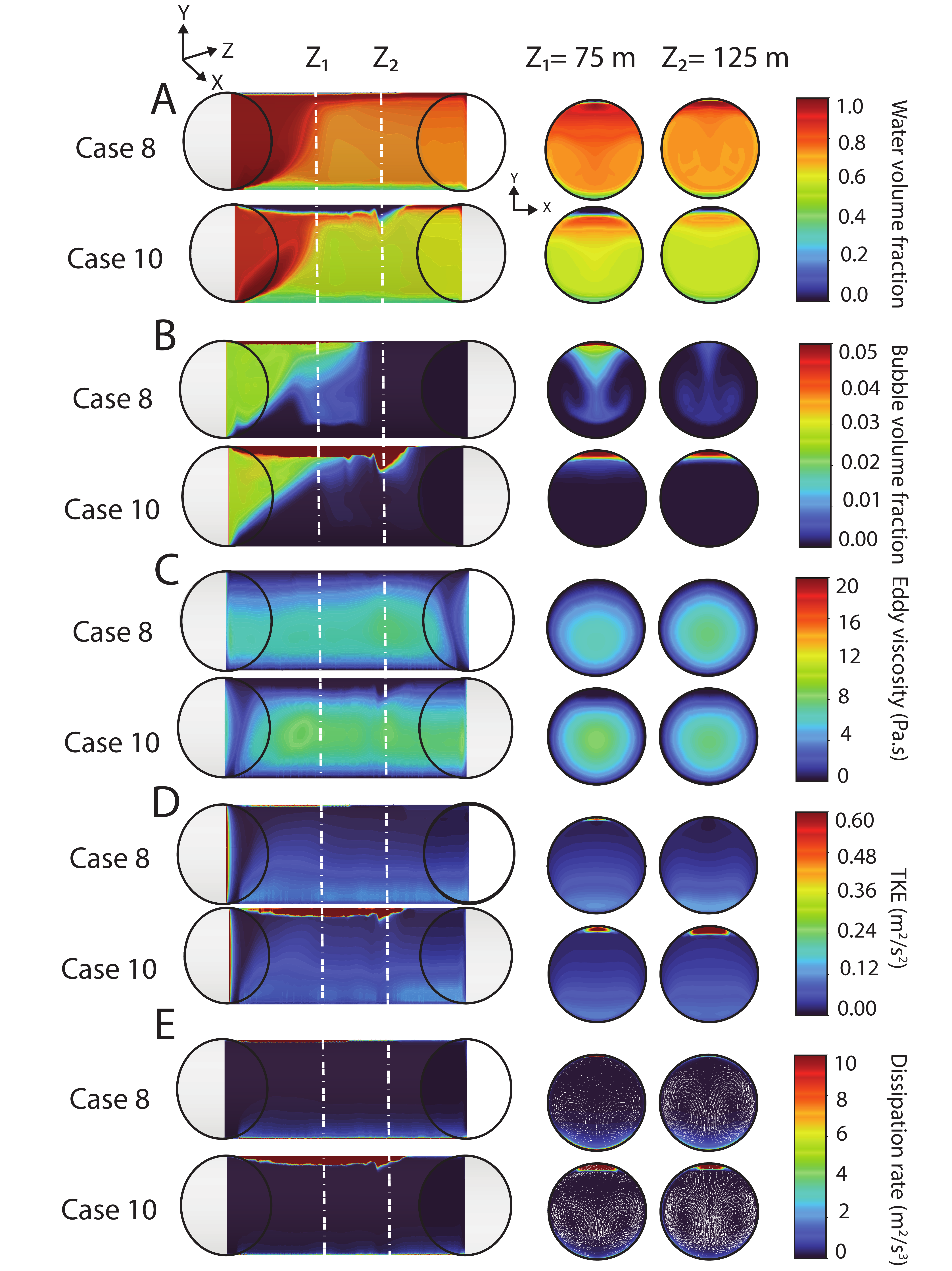} 
	\renewcommand{\captionfont}{\footnotesize}
	\caption{ The contour plots of (A) the water volume fraction,(B) bubble volume fraction, (C) eddy viscosity, (D) turbulent kinetic energy and (E) dissipation rate in the longitudinal section and two cross sections ($Z_1=75~m$, $Z_2=125~m$) for case 8 (bubble volume fraction 0.05 and bubble size 1000 ~$\mu m$) and case 10 (bubble volume fraction 0.20 and bubble size 1000 ~$\mu m$) at $t=60~s$ $ (t/T=0.2)$.}
	\label{fig:40}
\end{figure}

\subsection{Effect of bubbles fraction}

This section demonstrates the influence of bubble fraction on the flushing process.  Figure \ref{fig:40} A-E compares the water volume fraction and characteristics of turbulence in longitudinal section and two cross sections between flushing with and without bubbles and flushing with $5\%$ bubbles at $t=60~s$. Figure \ref{fig:40} A and B show the pipeline's water and bubble volume fraction. In case 10, the bubbles are found to float at the top of the pipeline, preventing water from flowing from the top region and moving through the solid bed. This results in a diluted bed, as shown in case 10. However, the bed is more resistant to removal because bubbles occupy larger regions of the crown of the pipeline. When bubbles are added to water, there is an increase in the volume fraction of water in the solid bed because of the increase in water distribution. Interestingly, in the longitudinal section, the ramp at the windward side becomes steeper. The results indicate that instead of carrying the particles downstream along the pipeline, the water flow is more likely to lift the particles vertically. It is also evident from the amount of water in the cross sections that the solid bed becomes more dispersed at the same $Z$ position for case 8. \\

\begin{figure*}[htbp] 
	\centering 
	\includegraphics[width=1\columnwidth] {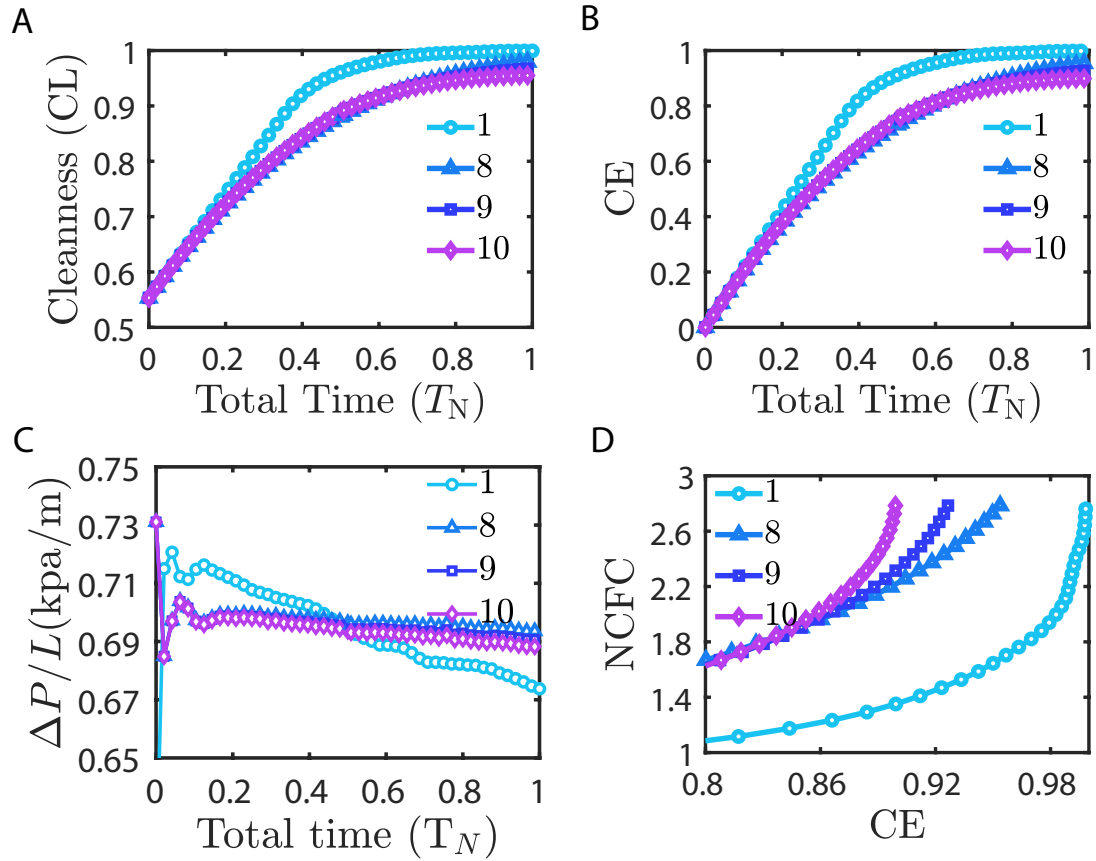}
	\renewcommand{\captionfont}{\footnotesize}
	\caption{(A) Cleanness, (B) cleanness efficiency, (C) pressure gradient over the pipeline, (D) normalized water consumption with respect to cleanness for case 1 (no bubble), 8 ($5\%$ bubble), 9 ($10\%$ bubble) and 10 ($20\%$ bubble).} 
	\label{fig:f20}
\end{figure*}

In addition, the shape of the bed suggests that the flushing process becomes slower with larger bubble volume fractions due to the eddy dissipation and turbulence suppression overcoming the bubble-solid particles interaction. Figure \ref{fig:40} C shows that the viscosity of the eddy increases when the volume fraction of the bubble increases, indicating the strength of the turbulence. From the Figure \ref{fig:40} B and C, it can be concluded that water flows through the solid bed as it flows through a porous structure, without suspending the particles and carrying them with the flow, while large volume of bubbles occupying the top region and have limited interaction with the solid bed.\\


As shown in Figure \ref{fig:40} D, the area of greatest turbulent kinetic energy moves from the lower region of the solid bed to the top of the pipeline, where bubbles accumulate. This indicates that these floating bubbles hinder the water flow. Figure \ref{fig:40} E shows that the dissipation rate of eddies is the largest in the bubble region, indicating that eddies becomes weaker when the flow reaches the solid bed compared to the case without bubbles. This is in accordance with the velocity vectors of the vortices shown in the cross sections at the windward side of the solid bed in Figure \ref{fig:40} E. These results suggest that adding bubbles to the water slows the flushing process~\cite{zakari2025dynamics}.


Figure \ref{fig:f20} describes the effect of the bubble fraction in the water on the overall cleaning performance. Figure \ref{fig:f20} A and B show that the time required to reach a specific CL or CE significantly increases after introducing bubbles, which is in accordance with the conclusion made from Figure \ref{fig:f20}. In addition, the volume fraction of the bubble has negligible effect on the flushing time. Figure \ref{fig:f20} C shows that after introducing bubbles, the pressure gradient in the pipeline reduces at a much lower rate compared to its counterpart without bubbles. This further shows that the water gradually flows through the solid bed instead of flushing particles downstream. Figure \ref{fig:f20} D shows that the water consumption at a specific CE almost doubles for the case with bubbles, and it gets higher as the bubble fraction increases. As a result, the CFD results recommend that the injection of bubbles during the flushing process of a pipeline could negatively affect process efficiency.\\

\section{Conclusion}
In this paper, a 3D mixture model coupled with KTGF to study the flushing process in a horizontal pipeline is presented. The model is validated with field data from industry and then the model is used to study the mechanisms of removing coarse particles whose sizes range from $161 ~\mu m$ to $672 ~\mu m$. The model is then used to study the effect of water velocity, bubble size, and bubble volume fractions on flushing performance and water consumption.

In this paper, it is demonstrated that the high velocity water removes particles in a stratified pattern. Smaller particles at the surface of the solid bed is first carried downstream by the water flow, revealing large particles at the bottom of the solid bed. Moreover, it is found that turbulence helps to suspend particles from the solid bed, which facilitates the flushing process. As a result, it is important to keep the water velocity above the critical velocity, at which the turbulence is strong enough to suspend largest particles in the solid bed to reach the most cost-efficient operating condition for flushing.

Furthermore, the role played by the bubbles in flushing process is explored. It is found that the small bubbles (50 ~$\mu m$ and 150 ~$\mu m$) enhance the flushing process due to their interaction with the solid content enhancing their suspension and lead to a faster and more effective flushing process. On the other hand the large bubbles (1000 ~$\mu m$) weakens the strength of turbulence before the flow reaches the solid bed, which worsen the flushing performance. The flow tends to lift the particles and flow through the solid bed rather than carrying the particles downstream. Moreover larger volumes of bubbles lead to greater eddy dissipation rates leading to weaker turbulence and solid content suspension and reducing the flushing efficiency when the dissipation effect overcomes the bubble interaction with solids. As a result, the model suggest to introduce low volume fraction (<5\%) of small bubbles in the water.

For all the cases studied, the amount of water consumption increases dramatically as the requirement of cleanness increases closer to unity. As a result, it is important to select proper target cleanness since pursuing an ultra clean pipeline can be costly. In conclusion, this model provides a guide for selecting the proper operating conditions for flushing process and provides insight in balancing the cost and flushing performance for the industry.

\section*{Supplementary data}

A supporting document is available on PSD details of model validation cases and the effect of temperature on the flushing process.

\section*{Data Availability Statement}
The data that support the findings of this study are available from the corresponding author upon reasonable request.

\section*{Acknowledgment}
The authors are thankful to the support from Canada Research Chairs Program, Discovery Project, Alliance Grant from Natural Sciences and Engineering Research Council of Canada (NSERC). The authors also acknowledge the Digital Research Alliance of Canada (https://alliancecan.ca/en) for continued support through regular access to high-performance computing resources, specifically the Cedar and Graham clusters.

\section*{Credit author statement}

M. Golchin: Formal analysis (lead); Investigation (lead); Methodology (equal); Software (lead); Validation (lead); Visualization (lead); Writing – review \& editing (equal). S. Chen: Writing – review \& editing (equal). S. Sharma: Writing – review \& editing (equal).  Y. Feng: Writing – review \& editing (equal). G. Shou: Industrial data (lead). P. Nikrityuk: Methodology (equal); Writing – review \& editing (supporting); S.G. Sontti: Conceptualization (equal); Methodology (equal); Visualization (equal), Writing – review \& editing (equal).; X. Zhang: Conceptualization (lead); Funding acquisition (lead); Methodology (lead); Project administration (lead); Resources (lead); Supervision (lead); Writing – review \& editing (equal). 

\section*{NOMENCLATURE}

\begin{longtable}{lp{12cm}}
	\hline
	\textbf{Symbol} & \textbf{Definition} \\
	\hline
	$A$ & Pipeline interface surface area (L$^2$) \\
	$D$ & Pipeline diameter (L) \\
	$R$ & Pipeline radius (L) \\
	$L$ & Pipeline Length (L) \\
	$d_p$ & Particle diameter (L) \\
	$C_v$ & Chord\textendash averaged concentration (--) \\
	$C_{v_{i}}$ & Local concentration (--) \\
	$C_{v_{s}}$ & Bulk concentration of sands (--) \\
	$g$ & Gravitational acceleration (L T$^{-2}$) \\
	$g_0$ & Radial distribution function (--) \\
	$p$ & Locally\textendash averaged pressure (M L$^{-1}$ T$^{-2}$) \\
	$t$ & Time (T) \\
	$T$ & Total time (T) \\
	$T_{N}$ & Total time (--) \\
	$v$ & Velocity (L T$^{-1}$) \\
	$u$ & Velocity (L T$^{-1}$) \\
	$u'$ & Velocity fluctuation (L T$^{-1}$) \\
	$V$ & Pipeline volume (L$^3$) \\
	$F$ & Force per unit volume (M L$^{-2}$ T$^{-2}$) \\
	$F_{l,si}$ & Interaction force of liquid and $i^\text{th}$ sand phase (M L$^{-2}$ T$^{-2}$) \\
	$C_\mathrm{fr}$ & Friction coefficient between solid phases (--) \\
	$x$ & Horizontal coordinate (L) \\
	$y$ & Vertical coordinate (L) \\
	$z$ & Axial coordinate (L) \\
	$\vec{v}_m$ & Mass-averaged (mixture) velocity (L T$^{-1}$) \\
	$\vec{v}_{k}$ & Velocity of phase $k$ (L T$^{-1}$) \\
	$\vec{v}_{pq}$ & Relative velocity between secondary phase $p$ and primary phase $q$ (L T$^{-1}$) \\
	$\vec{v}_{dr,k}$ & Drift velocity of phase $k$ (L T$^{-1}$) \\
	$\bar{v}_{pq}$ & Algebraic slip velocity (L T$^{-1}$) \\
	$\rho_k$ & Density of phase $k$ (M L$^{-3}$) \\
	$\rho_m$ & Mixture density (M L$^{-3}$) \\
	$\mu_k$ & Dynamic viscosity of phase $k$ (M L$^{-1}$ T$^{-1}$) \\
	$\mu_m$ & Mixture viscosity (M L$^{-1}$ T$^{-1}$) \\
	$\tau_p$ & Particle relaxation time (T) \\
	$\bar{a}$ & Particle acceleration (L T$^{-2}$) \\
	$c_k$ & Mass fraction of phase $k$ (--) \\
	$\alpha_k$ & Volume fraction of phase $k$ (--) \\
	$\alpha_s$ & Volume fraction of solid phase (--) \\
	$\vec{v}_{dr,s}$ & Drift velocity of the solid phase (L T$^{-1}$) \\
	$\mu_q$ & Viscosity of the continuous (primary) phase $q$ (M L$^{-1}$ T$^{-1}$) \\
	$\vec{v}_{p}$ & Velocity of secondary phase $p$ (L T$^{-1}$) \\
	$\vec{v}_{q}$ & Velocity of primary phase $q$ (L T$^{-1}$) \\
	$COR$ ($e$) & Coefficient of restitution (--) \\
	$k$ & Turbulent kinetic energy (L$^2$ T$^{-2}$) \\
	$I_{2D}$ & Second invariant of the deviatoric stress tensor (T$^{-1}$) \\
	$S$ & Turbulence source (M L$^{-1}$ T$^{-3}$) \\
	$CL$ & Cleanness (--) \\
	$CE$ & Cleanness efficiency (--) \\
	$DO_{i}$ & $i^\text{th}$ deposit occupation (--) \\
	$NDO$ & Normalized deposit occupation (--) \\
	$CFC_{c}^{n}$ & Carrier fluid consumption (L$^3$) \\
	$NCFC$ & Normalized carrier fluid consumption (--) \\
	$C_{s}$ & Solid content weight fraction (--) \\
	\hline
	\multicolumn{2}{l}{\textit{Greek symbols}} \\
	\hline
	$\alpha$ & Locally\textendash averaged volume fraction (--) \\
	$\beta$ & Coefficient of thermal expansion (K$^{-1}$) \\
	$\beta_{s}$ & Drag coefficient between granular material and carrier (--) \\
	$\mu$ & Dynamic viscosity (M L$^{-1}$ T$^{-1}$) \\
	$\mu_{l}$ & Dynamic viscosity of carrier (M L$^{-1}$ T$^{-1}$) \\
	$\mu_{b}$ & Bingham dynamic viscosity (M L$^{-1}$ T$^{-1}$) \\
	$\rho$ & Density (M L$^{-3}$) \\
	$\phi_{ls}$ & Energy exchange between the fluid and sand phases (M L$^2$ T$^{-2}$) \\
	$\gamma_{\Theta_{s}}$ & Collisional dissipation of energy (M L$^2$ T$^{-2}$) \\
	$\tau$ & Shear stress (M L$^{-1}$ T$^{-2}$) \\
	$\tau_{y}$ & Yield stress (M L$^{-1}$ T$^{-2}$) \\
	$\tau_{c}$ & Shear stress of granular coal phase (M L$^{-1}$ T$^{-2}$) \\
	$\tau_{s}$ & Shear stress of granular solid phase (M L$^{-1}$ T$^{-2}$) \\
	$\dot{\gamma}$ & Shear strain rate (T$^{-1}$) \\
	$\alpha_{s,\mathrm{max}}$ & Maximum packing limit (--) \\
	$\Theta$ & Granular temperature (L$^{-2}$ T$^{-2}$) \\
	$\theta$ & Temperature fluctuation (K) \\
	$\varphi$ & Angle of internal friction (--) \\
	$\phi$ & Volume fraction (--) \\
	$\mu_{t,m}$ & Turbulence viscosity (M L$^{-1}$ T$^{-1}$) \\
	$\epsilon$ & Dissipation rate (L$^2$ T$^{-3}$) \\
	$\lambda$ & Bulk viscosity (M L$^{-1}$ T$^{-1}$) \\
	$\rho_{m}$ & Mixture density (M L$^{-3}$) \\
	$\Omega_{k}$ & Angular velocity of pipeline rotation (T$^{-1}$) \\
	\hline
	\multicolumn{2}{l}{\textit{Subscripts}} \\
	\hline
	$c$ & Coal \\
	$b$ & Bubble \\
	$m$ & Mixture \\
	$si$ & Sand phase $i^\text{th}$ \\
	$ci$ & Coal phase $i^\text{th}$ \\
	$i$ & Phase or direction $i^\text{th}$ \\
	$j$ & Phase or direction $j^\text{th}$ \\
	$k$ & Direction $k^\text{th}$ \\
	$col$ & Collisional part of viscosity \\
	$kin$ & Kinetic part of viscosity \\
	$fr$ & Frictional part of viscosity \\
	$td$ & Turbulent dispersion \\
	$drag$ & Drag \\
	$0$ & Initial \\
	$cf$ & Carrier fluid \\
	$cv$ & Control volume \\
	\hline
	\multicolumn{2}{l}{\textit{Abbreviations}} \\
	\hline
	$PSD$ & Particle size distribution (--) \\
	$RSM$ & Reynolds Stress Model (--) \\
	$RANS$ & Reynolds-Averaged Navier-Stokes (--) \\
	\hline
\end{longtable}

\newpage
\bibliography{ref}

\begin{thebibliography}{70}
\expandafter\ifx\csname natexlab\endcsname\relax\def\natexlab#1{#1}\fi
\providecommand{\url}[1]{\texttt{#1}}
\providecommand{\href}[2]{#2}
\providecommand{\path}[1]{#1}
\providecommand{\DOIprefix}{doi:}
\providecommand{\ArXivprefix}{arXiv:}
\providecommand{\URLprefix}{URL: }
\providecommand{\Pubmedprefix}{pmid:}
\providecommand{\doi}[1]{\href{http://dx.doi.org/#1}{\path{#1}}}
\providecommand{\Pubmed}[1]{\href{pmid:#1}{\path{#1}}}
\providecommand{\bibinfo}[2]{#2}
\ifx\xfnm\relax \def\xfnm[#1]{\unskip,\space#1}\fi
\bibitem[{Pullum et~al.(2018)Pullum, Boger, and Sofra}]{intro1pullum}
\bibinfo{author}{L.~Pullum}, \bibinfo{author}{D.~V. Boger},
  \bibinfo{author}{F.~Sofra},
\newblock \bibinfo{title}{Hydraulic mineral waste transport and storage},
\newblock \bibinfo{journal}{Annu. Rev. Fluid Mech.} \bibinfo{volume}{50}
  (\bibinfo{year}{2018}) \bibinfo{pages}{157--185}.
\bibitem[{Mohaibes and Heinonen-Tanski(2004)}]{intro2Mohaibes}
\bibinfo{author}{M.~Mohaibes}, \bibinfo{author}{H.~Heinonen-Tanski},
\newblock \bibinfo{title}{Aerobic thermophilic treatment of farm slurry and
  food wastes},
\newblock \bibinfo{journal}{Bioresour. Technol.} \bibinfo{volume}{95}
  (\bibinfo{year}{2004}) \bibinfo{pages}{245--254}.
\bibitem[{Li et~al.(2018)Li, He, Liu, and Huang}]{intro3li}
\bibinfo{author}{M.~Li}, \bibinfo{author}{Y.~He}, \bibinfo{author}{Y.~Liu},
  \bibinfo{author}{C.~Huang},
\newblock \bibinfo{title}{Effect of interaction of particles with different
  sizes on particle kinetics in multi-sized slurry transport by pipeline},
\newblock \bibinfo{journal}{Powder Technol.} \bibinfo{volume}{338}
  (\bibinfo{year}{2018}) \bibinfo{pages}{915--930}.
\bibitem[{Gillies et~al.(1991)Gillies, Shook, and Wilson}]{gilles1}
\bibinfo{author}{R.~Gillies}, \bibinfo{author}{C.~Shook},
  \bibinfo{author}{K.~Wilson},
\newblock \bibinfo{title}{An improved two layer model for horizontal slurry
  pipeline flow},
\newblock \bibinfo{journal}{Can. J. Chem. Eng.} \bibinfo{volume}{69}
  (\bibinfo{year}{1991}) \bibinfo{pages}{173--178}.
\bibitem[{Hosseini et~al.(2024)Hosseini, Mozaffarian, Dabir, and Van~den
  Akker}]{hosseini2024coupled}
\bibinfo{author}{S.~F. Hosseini}, \bibinfo{author}{M.~Mozaffarian},
  \bibinfo{author}{B.~Dabir}, \bibinfo{author}{H.~E. Van~den Akker},
\newblock \bibinfo{title}{A coupled dem-cfd analysis of asphaltene particles
  agglomeration in turbulent pipe flow},
\newblock \bibinfo{journal}{Chem. Eng. J.} \bibinfo{volume}{486}
  (\bibinfo{year}{2024}) \bibinfo{pages}{150070}.
\bibitem[{Leporini et~al.(2019)Leporini, Marchetti, Corvaro, di~Giovine,
  Polonara, and Terenzi}]{blockadeintrolep}
\bibinfo{author}{M.~Leporini}, \bibinfo{author}{B.~Marchetti},
  \bibinfo{author}{F.~Corvaro}, \bibinfo{author}{G.~di~Giovine},
  \bibinfo{author}{F.~Polonara}, \bibinfo{author}{A.~Terenzi},
\newblock \bibinfo{title}{Sand transport in multiphase flow mixtures in a
  horizontal pipeline: An experimental investigation},
\newblock \bibinfo{journal}{Petroleum} \bibinfo{volume}{5}
  (\bibinfo{year}{2019}) \bibinfo{pages}{161--170}.
\bibitem[{Scott and Yi(1999)}]{blockade2scott}
\bibinfo{author}{S.~Scott}, \bibinfo{author}{J.~Yi},
\newblock \bibinfo{title}{Flow testing methods to detect and characterize
  partial blockages in looped subsea flowlines},
\newblock \bibinfo{journal}{J. Energy Resour. Technol. Sep}
  (\bibinfo{year}{1999}).
\bibitem[{Lai and Shen(1996)}]{flushingprocess}
\bibinfo{author}{J.-S. Lai}, \bibinfo{author}{H.~W. Shen},
\newblock \bibinfo{title}{Flushing sediment through reservoirs},
\newblock \bibinfo{journal}{J. Hydraul. Res.} \bibinfo{volume}{34}
  (\bibinfo{year}{1996}) \bibinfo{pages}{237--255}.
\bibitem[{Li and Chen(2022)}]{li2022multi}
\bibinfo{author}{J.~Li}, \bibinfo{author}{X.~Chen},
\newblock \bibinfo{title}{A multi-dimensional two-phase mixture model for
  intense sediment transport in sheet flow and around pipeline},
\newblock \bibinfo{journal}{Phys. Fluids.} \bibinfo{volume}{34}
  (\bibinfo{year}{2022}).
\bibitem[{Camenen and Larson(2008)}]{camenen2008general}
\bibinfo{author}{B.~Camenen}, \bibinfo{author}{M.~Larson},
\newblock \bibinfo{title}{A general formula for noncohesive suspended sediment
  transport},
\newblock \bibinfo{journal}{J. Coast. Res.} \bibinfo{volume}{24}
  (\bibinfo{year}{2008}) \bibinfo{pages}{615--627}.
\bibitem[{Zhang et~al.(2023)Zhang, Li, and Ma}]{DEM_mono2023}
\bibinfo{author}{S.~Zhang}, \bibinfo{author}{B.~Li}, \bibinfo{author}{H.~Ma},
\newblock \bibinfo{title}{Numerical investigation of scour around the monopile
  using cfd-dem coupling method},
\newblock \bibinfo{journal}{Coast. Eng.} \bibinfo{volume}{183}
  (\bibinfo{year}{2023}) \bibinfo{pages}{104334}.
\bibitem[{Ma and Li(2023)}]{CFDCGDEM}
\bibinfo{author}{H.~Ma}, \bibinfo{author}{B.~Li},
\newblock \bibinfo{title}{Cfd-cgdem coupling model for scour process simulation
  of submarine pipelines},
\newblock \bibinfo{journal}{Ocean Eng.} \bibinfo{volume}{271}
  (\bibinfo{year}{2023}) \bibinfo{pages}{113789}.
\bibitem[{Zhao(2022)}]{zhao2022review}
\bibinfo{author}{M.~Zhao},
\newblock \bibinfo{title}{A review on recent development of numerical modelling
  of local scour around hydraulic and marine structures},
\newblock \bibinfo{journal}{J. Mar. Sci. Eng} \bibinfo{volume}{10}
  (\bibinfo{year}{2022}) \bibinfo{pages}{1139}.
\bibitem[{Ofei and Ismail(2016)}]{ofei2016eulerian}
\bibinfo{author}{T.~N. Ofei}, \bibinfo{author}{A.~Y. Ismail},
\newblock \bibinfo{title}{Eulerian-eulerian simulation of particle-liquid
  slurry flow in horizontal pipe},
\newblock \bibinfo{journal}{J. Pet. Eng.} \bibinfo{volume}{2016}
  (\bibinfo{year}{2016}) \bibinfo{pages}{5743471}.
\bibitem[{Thakur et~al.(2022)Thakur, Kumar, Banerjee, Chaudhari, and
  Gaurav}]{thakur2022hydrodynamic}
\bibinfo{author}{A.~K. Thakur}, \bibinfo{author}{R.~Kumar},
  \bibinfo{author}{N.~Banerjee}, \bibinfo{author}{P.~Chaudhari},
  \bibinfo{author}{G.~K. Gaurav},
\newblock \bibinfo{title}{Hydrodynamic modeling of liquid-solid flow in
  polyolefin slurry reactors using cfd techniques--a critical analysis},
\newblock \bibinfo{journal}{Powder Technol.} \bibinfo{volume}{405}
  (\bibinfo{year}{2022}) \bibinfo{pages}{117544}.
\bibitem[{Messa et~al.(2021)Messa, Yang, Adedeji, Ch{\'a}ra, Duarte,
  Matou{\v{s}}ek, Rasteiro, Sanders, Silva, and
  de~Souza}]{messa2021computational}
\bibinfo{author}{G.~V. Messa}, \bibinfo{author}{Q.~Yang},
  \bibinfo{author}{O.~E. Adedeji}, \bibinfo{author}{Z.~Ch{\'a}ra},
  \bibinfo{author}{C.~A.~R. Duarte}, \bibinfo{author}{V.~Matou{\v{s}}ek},
  \bibinfo{author}{M.~G. Rasteiro}, \bibinfo{author}{R.~S. Sanders},
  \bibinfo{author}{R.~C. Silva}, \bibinfo{author}{F.~J. de~Souza},
\newblock \bibinfo{title}{Computational fluid dynamics modelling of
  liquid--solid slurry flows in pipelines: State-of-the-art and future
  perspectives},
\newblock \bibinfo{journal}{Processes} \bibinfo{volume}{9}
  (\bibinfo{year}{2021}) \bibinfo{pages}{1566}.
\bibitem[{Coverston(2019)}]{E-EI2}
\bibinfo{author}{J.~S. Coverston}, \bibinfo{title}{Numerical Simulation of
  Flushing Deposits in Pipelines}, Master's thesis, Florida International
  University, \bibinfo{year}{2019}.
\bibitem[{Alihosseini and Thamsen(2019)}]{sewer_DEM2}
\bibinfo{author}{M.~Alihosseini}, \bibinfo{author}{P.~U. Thamsen},
\newblock \bibinfo{title}{On scouring efficiency of flush waves in sewers: A
  numerical and experimental study},
\newblock in: \bibinfo{booktitle}{Fluids Engineering Division Summer Meeting},
  volume \bibinfo{volume}{59087}, \bibinfo{organization}{American Society of
  Mechanical Engineers}, \bibinfo{year}{2019}, p. \bibinfo{pages}{V005T05A080}.
\bibitem[{Shi et~al.(2021)Shi, Yuan, and Li}]{swirl-shi}
\bibinfo{author}{H.~Shi}, \bibinfo{author}{J.~Yuan}, \bibinfo{author}{Y.~Li},
\newblock \bibinfo{title}{The impact of swirls on slurry flows in horizontal
  pipelines},
\newblock \bibinfo{journal}{J. Mar. Sci. Eng.} \bibinfo{volume}{9}
  (\bibinfo{year}{2021}) \bibinfo{pages}{1201}.
\bibitem[{Janu{\'a}rio and Maia(2020)}]{januario2020cfd}
\bibinfo{author}{J.~Janu{\'a}rio}, \bibinfo{author}{C.~Maia},
\newblock \bibinfo{title}{Cfd-dem simulation to predict the critical velocity
  of slurry flows},
\newblock \bibinfo{journal}{J. Appl. Fluid Mech.} \bibinfo{volume}{13}
  (\bibinfo{year}{2020}) \bibinfo{pages}{161--168}.
\bibitem[{Yuan et~al.(2022)Yuan, Chen, and Ding}]{airbubbleflush}
\bibinfo{author}{W.~Yuan}, \bibinfo{author}{F.~Chen},
  \bibinfo{author}{J.~Ding},
\newblock \bibinfo{title}{Practical analysis of cleaning water supply pipeline
  using air and water flushing technology},
\newblock in: \bibinfo{booktitle}{IOP Conference Series: Earth and
  Environmental Science}, \bibinfo{number}{1}, \bibinfo{organization}{IOP
  Publishing}, \bibinfo{year}{2022}, p. \bibinfo{pages}{012088}.
\bibitem[{Bai et~al.(2024)Bai, Wang, Wu, Li, Wang, Chen, and
  Liu}]{bai2024effect}
\bibinfo{author}{L.~Bai}, \bibinfo{author}{H.~Wang}, \bibinfo{author}{A.~Wu},
  \bibinfo{author}{H.~Li}, \bibinfo{author}{S.~Wang},
  \bibinfo{author}{C.~Chen}, \bibinfo{author}{C.~Liu},
\newblock \bibinfo{title}{Effect of compressed gas on deep mine long-distance
  backfill pipe flushing: From laboratory to industrial tests},
\newblock \bibinfo{journal}{Adv. Civ. Eng. Mater.} \bibinfo{volume}{2024}
  (\bibinfo{year}{2024}) \bibinfo{pages}{6211202}.
\bibitem[{Motamed~Dashliborun et~al.(2020)Motamed~Dashliborun, Zhou, Esmaeili,
  and Zhang}]{motamed2020microbubble}
\bibinfo{author}{A.~Motamed~Dashliborun}, \bibinfo{author}{J.~Zhou},
  \bibinfo{author}{P.~Esmaeili}, \bibinfo{author}{X.~Zhang},
\newblock \bibinfo{title}{Microbubble-enhanced recovery of residual bitumen
  from the tailings of oil sands extraction in a laboratory-scale pipeline},
\newblock \bibinfo{journal}{Energy \& Fuels} \bibinfo{volume}{34}
  (\bibinfo{year}{2020}) \bibinfo{pages}{16476--16485}.
\bibitem[{Gao et~al.(2021)Gao, Dashliborun, Zhou, and Zhang}]{gao2021formation}
\bibinfo{author}{Y.~Gao}, \bibinfo{author}{A.~M. Dashliborun},
  \bibinfo{author}{J.~Z. Zhou}, \bibinfo{author}{X.~Zhang},
\newblock \bibinfo{title}{Formation and stability of cavitation microbubbles in
  process water from the oilsands industry},
\newblock \bibinfo{journal}{Ind. Eng. Chem. Res.} \bibinfo{volume}{60}
  (\bibinfo{year}{2021}) \bibinfo{pages}{3198--3209}.
\bibitem[{Zhou et~al.(2022)Zhou, Sontti, Zhou, Esmaeili, and
  Zhang}]{zhou2022microbubble}
\bibinfo{author}{K.~Zhou}, \bibinfo{author}{S.~G. Sontti},
  \bibinfo{author}{J.~Zhou}, \bibinfo{author}{P.~Esmaeili},
  \bibinfo{author}{X.~Zhang},
\newblock \bibinfo{title}{Microbubble-enhanced bitumen separation from tailing
  slurries with high solid contents},
\newblock \bibinfo{journal}{Ind. Eng. Chem. Res.} \bibinfo{volume}{61}
  (\bibinfo{year}{2022}) \bibinfo{pages}{17327--17341}.
\bibitem[{Zhou et~al.(2025)Zhou, Sharma, Sontti, and
  Zhang}]{zhou2025experimental}
\bibinfo{author}{K.~Zhou}, \bibinfo{author}{S.~Sharma}, \bibinfo{author}{S.~G.
  Sontti}, \bibinfo{author}{X.~Zhang},
\newblock \bibinfo{title}{Experimental and numerical study of
  microbubble-enhanced separation of aged microplastics in a slurry flow},
\newblock \bibinfo{journal}{Sep. Purif. Technol.} \bibinfo{volume}{359}
  (\bibinfo{year}{2025}) \bibinfo{pages}{130298}.
\bibitem[{Sadeghi et~al.(2023)Sadeghi, Sontti, Zheng, and Zhang}]{E-EI9}
\bibinfo{author}{M.~Sadeghi}, \bibinfo{author}{S.~G. Sontti},
  \bibinfo{author}{E.~Zheng}, \bibinfo{author}{X.~Zhang},
\newblock \bibinfo{title}{Computational fluid dynamics (cfd) simulation of
  three--phase non--newtonian slurry flows in industrial horizontal pipelines},
\newblock \bibinfo{journal}{Chem. Eng. Sci.} \bibinfo{volume}{270}
  (\bibinfo{year}{2023}) \bibinfo{pages}{118513}.
\bibitem[{Sadeghi et~al.(2022)Sadeghi, Li, Zheng, Sontti, Esmaeili, and
  Zhang}]{SCFD}
\bibinfo{author}{M.~Sadeghi}, \bibinfo{author}{S.~Li},
  \bibinfo{author}{E.~Zheng}, \bibinfo{author}{S.~G. Sontti},
  \bibinfo{author}{P.~Esmaeili}, \bibinfo{author}{X.~Zhang},
\newblock \bibinfo{title}{Cfd simulation of turbulent non-newtonian slurry
  flows in horizontal pipelines},
\newblock \bibinfo{journal}{Ind. Eng. Chem. Res.} \bibinfo{volume}{61}
  (\bibinfo{year}{2022}) \bibinfo{pages}{5324--5339}.
\bibitem[{Sontti et~al.(2023)Sontti, Sadeghi, Zhou, Zheng, and
  Zhang}]{Sekhar2023POF}
\bibinfo{author}{S.~G. Sontti}, \bibinfo{author}{M.~Sadeghi},
  \bibinfo{author}{K.~Zhou}, \bibinfo{author}{E.~Zheng},
  \bibinfo{author}{X.~Zhang},
\newblock \bibinfo{title}{Computational fluid dynamics investigation of bitumen
  residues in oil sands tailings transport in an industrial horizontal pipe},
\newblock \bibinfo{journal}{Phys. Fluids} \bibinfo{volume}{35}
  (\bibinfo{year}{2023}).
\bibitem[{Sontti and Zhang(2023)}]{sontti2023numerical}
\bibinfo{author}{S.~G. Sontti}, \bibinfo{author}{X.~Zhang},
\newblock \bibinfo{title}{Numerical insights from a population balance model
  into the distribution of bitumen residues in industrial horizontal pipes
  during the hydrotransport of oil sands tailings},
\newblock \bibinfo{journal}{Ind. Eng. Chem. Res.} \bibinfo{volume}{63}
  (\bibinfo{year}{2023}) \bibinfo{pages}{691--705}.
\bibitem[{Fluent(2020)}]{Ansys}
Fluent, \bibinfo{title}{A. Ansys fluent 20.0 user’s guide, ANSYS FLUENT Inc.,
  2020.}, \bibinfo{year}{2020}.
\bibitem[{Liu et~al.(2022{\natexlab{a}})Liu, He, Li, Huang, and Liu}]{drag1}
\bibinfo{author}{W.~Liu}, \bibinfo{author}{Y.~He}, \bibinfo{author}{M.~Li},
  \bibinfo{author}{C.~Huang}, \bibinfo{author}{Y.~Liu},
\newblock \bibinfo{title}{Effect of drag models on hydrodynamic behaviors of
  slurry flows in horizontal pipes},
\newblock \bibinfo{journal}{Phys. Fluids} \bibinfo{volume}{34}
  (\bibinfo{year}{2022}{\natexlab{a}}).
\bibitem[{Liu et~al.(2022{\natexlab{b}})Liu, He, Li, Liu, and Huang}]{drag3}
\bibinfo{author}{W.~Liu}, \bibinfo{author}{Y.~He}, \bibinfo{author}{M.~Li},
  \bibinfo{author}{Y.~Liu}, \bibinfo{author}{C.~Huang},
\newblock \bibinfo{title}{Effect of specularity coefficient on hydrodynamic
  behaviors of slurry flows in horizontal pipes},
\newblock \bibinfo{journal}{Ocean Eng.} \bibinfo{volume}{246}
  (\bibinfo{year}{2022}{\natexlab{b}}) \bibinfo{pages}{110617}.
\bibitem[{Lun et~al.(1984)Lun, Savage, Jeffrey, and Chepurniy}]{lun}
\bibinfo{author}{C.~K. Lun}, \bibinfo{author}{S.~B. Savage},
  \bibinfo{author}{D.~Jeffrey}, \bibinfo{author}{N.~Chepurniy},
\newblock \bibinfo{title}{Kinetic theories for granular flow: inelastic
  particles in couette flow and slightly inelastic particles in a general
  flowfield},
\newblock \bibinfo{journal}{J. Fluid Mech.} \bibinfo{volume}{140}
  (\bibinfo{year}{1984}) \bibinfo{pages}{223--256}.
\bibitem[{Gidaspow(1994)}]{gidaspowbook}
\bibinfo{author}{D.~Gidaspow}, \bibinfo{title}{Multiphase flow and
  fluidization: continuum and kinetic theory descriptions},
  \bibinfo{publisher}{Academic press}, \bibinfo{year}{1994}.
\bibitem[{Gidaspow et~al.(1991)Gidaspow, Bezburuah, and Ding}]{gidaspow2}
\bibinfo{author}{D.~Gidaspow}, \bibinfo{author}{R.~Bezburuah},
  \bibinfo{author}{J.~Ding}, \bibinfo{title}{Hydrodynamics of circulating
  fluidized beds: kinetic theory approach}, \bibinfo{type}{Technical Report},
  Illinois Inst. of Tech., Chicago, IL (United States). Dept. of Chemical~…,
  \bibinfo{year}{1991}.
\bibitem[{Koutsourakis et~al.(2012)Koutsourakis, Bartzis, and
  Markatos}]{turbu2}
\bibinfo{author}{N.~Koutsourakis}, \bibinfo{author}{J.~G. Bartzis},
  \bibinfo{author}{N.~C. Markatos},
\newblock \bibinfo{title}{Evaluation of reynolds stress, k-$\varepsilon$ and
  rng k-$\varepsilon$ turbulence models in street canyon flows using various
  experimental datasets},
\newblock \bibinfo{journal}{Environ. Fluid Mech.} \bibinfo{volume}{12}
  (\bibinfo{year}{2012}) \bibinfo{pages}{379--403}.
\bibitem[{Huo et~al.(2024)Huo, Golchin, Zhou, Abraham, Sontti, and
  Zhang}]{yiyi}
\bibinfo{author}{Y.~Huo}, \bibinfo{author}{M.~Golchin},
  \bibinfo{author}{K.~Zhou}, \bibinfo{author}{A.~Abraham},
  \bibinfo{author}{S.~G. Sontti}, \bibinfo{author}{X.~Zhang},
\newblock \bibinfo{title}{Effects of coal particles on microbubble-enhanced
  bitumen separation in the concentrated slurry flow of oil sands tailings},
\newblock \bibinfo{journal}{Ind. Eng. Chem. Res.}  (\bibinfo{year}{2024}).
\bibitem[{Launder et~al.(1975)Launder, Reece, and Rodi}]{turbu1}
\bibinfo{author}{B.~E. Launder}, \bibinfo{author}{G.~J. Reece},
  \bibinfo{author}{W.~Rodi},
\newblock \bibinfo{title}{Progress in the development of a reynolds-stress
  turbulence closure},
\newblock \bibinfo{journal}{J. Fluid Mech.} \bibinfo{volume}{68}
  (\bibinfo{year}{1975}) \bibinfo{pages}{537--566}.
\bibitem[{Durbin(1993)}]{turbu3}
\bibinfo{author}{P.~Durbin},
\newblock \bibinfo{title}{A reynolds stress model for near-wall turbulence},
\newblock \bibinfo{journal}{J. Fluid Mech.} \bibinfo{volume}{249}
  (\bibinfo{year}{1993}) \bibinfo{pages}{465--498}.
\bibitem[{Hanjali{\'c} and Launder(1972)}]{turbu4}
\bibinfo{author}{K.~Hanjali{\'c}}, \bibinfo{author}{B.~E. Launder},
\newblock \bibinfo{title}{A reynolds stress model of turbulence and its
  application to thin shear flows},
\newblock \bibinfo{journal}{J. Fluid Mech.} \bibinfo{volume}{52}
  (\bibinfo{year}{1972}) \bibinfo{pages}{609--638}.
\bibitem[{Hinze(1975)}]{turbu5}
\bibinfo{author}{J.~Hinze}, \bibinfo{title}{Turbulence},
  \bibinfo{publisher}{McGraw-Hill}, \bibinfo{year}{1975}.
\bibitem[{Baker et~al.(2019)Baker, Johnson, Flynn, Hemida, Quinn, Soper, and
  Sterling}]{baker2019train}
\bibinfo{author}{C.~Baker}, \bibinfo{author}{T.~Johnson},
  \bibinfo{author}{D.~Flynn}, \bibinfo{author}{H.~Hemida},
  \bibinfo{author}{A.~Quinn}, \bibinfo{author}{D.~Soper},
  \bibinfo{author}{M.~Sterling}, \bibinfo{title}{Train aerodynamics:
  fundamentals and applications}, \bibinfo{publisher}{Butterworth-Heinemann},
  \bibinfo{year}{2019}.
\bibitem[{Abdulrahman(2022)}]{abdulbubble}
\bibinfo{author}{M.~W. Abdulrahman},
\newblock \bibinfo{title}{Temperature profiles of a direct contact heat
  transfer in a slurry bubble column},
\newblock \bibinfo{journal}{Chem. Eng. Res. Des.} \bibinfo{volume}{182}
  (\bibinfo{year}{2022}) \bibinfo{pages}{183--193}.
\bibitem[{Ma et~al.(2022)Ma, Liu, Tang, Liu, Yang, and Yang}]{elbowiceslurry}
\bibinfo{author}{K.~Ma}, \bibinfo{author}{Z.~Liu}, \bibinfo{author}{Y.~Tang},
  \bibinfo{author}{X.~Liu}, \bibinfo{author}{Y.~Yang},
  \bibinfo{author}{S.~Yang},
\newblock \bibinfo{title}{Numerical investigation on ice slurry flow in
  horizontal elbow pipes},
\newblock \bibinfo{journal}{Therm. Sci. Eng. Prog} \bibinfo{volume}{27}
  (\bibinfo{year}{2022}) \bibinfo{pages}{101083}.
\bibitem[{Hu and Tao(2022)}]{cicepigging}
\bibinfo{author}{J.~Hu}, \bibinfo{author}{T.~Tao},
\newblock \bibinfo{title}{Numerical investigation of ice pigging isothermal
  flow in water-supply pipelines cleaning},
\newblock \bibinfo{journal}{Chem. Eng. Res. Des.} \bibinfo{volume}{182}
  (\bibinfo{year}{2022}) \bibinfo{pages}{428--437}.
\bibitem[{Spalding(2015)}]{launder}
\bibinfo{author}{D.~B. Spalding}, \bibinfo{title}{Numerical prediction of flow,
  heat transfer, turbulence and combustion}, \bibinfo{publisher}{Elsevier},
  \bibinfo{year}{2015}.
\bibitem[{Thakur et~al.(2022)Thakur, Kumar, Banerjee, Chaudhari, and
  Gaurav}]{cGauravslurrypowder}
\bibinfo{author}{A.~K. Thakur}, \bibinfo{author}{R.~Kumar},
  \bibinfo{author}{N.~Banerjee}, \bibinfo{author}{P.~Chaudhari},
  \bibinfo{author}{G.~K. Gaurav},
\newblock \bibinfo{title}{Hydrodynamic modeling of liquid-solid flow in
  polyolefin slurry reactors using cfd techniques--a critical analysis},
\newblock \bibinfo{journal}{Powder Technol.} \bibinfo{volume}{405}
  (\bibinfo{year}{2022}) \bibinfo{pages}{117544}.
\bibitem[{Ro and Ryou(2012)}]{kepsilonturb}
\bibinfo{author}{K.~Ro}, \bibinfo{author}{H.~Ryou},
\newblock \bibinfo{title}{Development of the modified $k$--$\varepsilon$ turbulence model of power-law fluid for engineering applications},
\newblock \bibinfo{journal}{Sci. China Technol. Sci.} \bibinfo{volume}{55}
  (\bibinfo{year}{2012}) \bibinfo{pages}{276--284}.
\bibitem[{Ting et~al.(2019)Ting, Miedema, and Xiuhan}]{Slurryocean}
\bibinfo{author}{X.~Ting}, \bibinfo{author}{S.~A. Miedema},
  \bibinfo{author}{C.~Xiuhan},
\newblock \bibinfo{title}{Comparative analysis between cfd model and dhlldv
  model in fully-suspended slurry flow},
\newblock \bibinfo{journal}{Ocean Eng.} \bibinfo{volume}{181}
  (\bibinfo{year}{2019}) \bibinfo{pages}{29--42}.
\bibitem[{Katsamis et~al.(2022)Katsamis, Craft, Iacovides, and Uribe}]{Rans23d}
\bibinfo{author}{C.~Katsamis}, \bibinfo{author}{T.~Craft},
  \bibinfo{author}{H.~Iacovides}, \bibinfo{author}{J.~C. Uribe},
\newblock \bibinfo{title}{Use of 2-d and 3-d unsteady rans in the computation
  of wall bounded buoyant flows},
\newblock \bibinfo{journal}{Int. J. Heat Fluid Flow} \bibinfo{volume}{93}
  (\bibinfo{year}{2022}) \bibinfo{pages}{108914}.
\bibitem[{Liu et~al.(2022)Liu, He, Li, Huang, and Liu}]{wliuhozslurry}
\bibinfo{author}{W.~Liu}, \bibinfo{author}{Y.~He}, \bibinfo{author}{M.~Li},
  \bibinfo{author}{C.~Huang}, \bibinfo{author}{Y.~Liu},
\newblock \bibinfo{title}{Effect of drag models on hydrodynamic behaviors of
  slurry flows in horizontal pipes},
\newblock \bibinfo{journal}{Phys. Fluids} \bibinfo{volume}{34}
  (\bibinfo{year}{2022}).
\bibitem[{Spalding(2015)}]{yplus1}
\bibinfo{author}{D.~B. Spalding}, \bibinfo{title}{Numerical prediction of flow,
  heat transfer, turbulence and combustion}, \bibinfo{publisher}{Elsevier},
  \bibinfo{year}{2015}.
\bibitem[{Launder and Spalding(1983)}]{yplus2}
\bibinfo{author}{B.~E. Launder}, \bibinfo{author}{D.~B. Spalding},
\newblock \bibinfo{title}{The numerical computation of turbulent flows},
\newblock in: \bibinfo{booktitle}{Numerical prediction of flow, heat transfer,
  turbulence and combustion}, \bibinfo{publisher}{Elsevier},
  \bibinfo{year}{1983}, pp. \bibinfo{pages}{96--116}.
\bibitem[{Zengeni(2016)}]{zengeni-bingham}
\bibinfo{author}{B.~T. Zengeni}, \bibinfo{title}{Bingham yield stress and
  Bingham plastic viscosity of homogeneous Non-Newtonian slurries}, Ph.D.
  thesis, Cape Peninsula University of Technology, \bibinfo{year}{2016}.
\bibitem[{Brown and Heywood(1991)}]{fullydevelopement}
\bibinfo{author}{N.~P. Brown}, \bibinfo{author}{N.~I. Heywood},
  \bibinfo{title}{Slurry Handling: Design of solid-liquid systems},
  \bibinfo{publisher}{Springer Science \& Business Media},
  \bibinfo{year}{1991}.
\bibitem[{Burns et~al.(2004)Burns, Frank, Hamill, Shi et~al.}]{burns2004favre}
\bibinfo{author}{A.~D. Burns}, \bibinfo{author}{T.~Frank},
  \bibinfo{author}{I.~Hamill}, \bibinfo{author}{J.-M. Shi}, et~al.,
\newblock \bibinfo{title}{The favre averaged drag model for turbulent
  dispersion in eulerian multi-phase flows},
\newblock in: \bibinfo{booktitle}{5th international conference on multiphase
  flow, ICMF}, volume~\bibinfo{volume}{4}, \bibinfo{organization}{ICMF},
  \bibinfo{year}{2004}, pp. \bibinfo{pages}{1--17}.
\bibitem[{Zhang et~al.(2021)Zhang, Nathan, Tian, and Chin}]{ppcollision}
\bibinfo{author}{X.~Zhang}, \bibinfo{author}{G.~J. Nathan},
  \bibinfo{author}{Z.~F. Tian}, \bibinfo{author}{R.~C. Chin},
\newblock \bibinfo{title}{The influence of the coefficient of restitution on
  flow regimes within horizontal particle-laden pipe flows},
\newblock \bibinfo{journal}{Phys. Fluids} \bibinfo{volume}{33}
  (\bibinfo{year}{2021}).
\bibitem[{Peker and Helvaci(2011)}]{peker2011solid}
\bibinfo{author}{S.~M. Peker}, \bibinfo{author}{S.~S. Helvaci},
  \bibinfo{title}{Solid-liquid two phase flow}, \bibinfo{publisher}{Elsevier},
  \bibinfo{year}{2011}.
\bibitem[{Wang et~al.(2022)Wang, Wang, Chen, He, Chen, and
  Zhang}]{wang2022dynamic}
\bibinfo{author}{C.~Wang}, \bibinfo{author}{F.~Wang},
  \bibinfo{author}{W.~Chen}, \bibinfo{author}{Q.~He},
  \bibinfo{author}{X.~Chen}, \bibinfo{author}{Z.~Zhang},
\newblock \bibinfo{title}{A dynamic particle scale-driven interphase force
  model for water-sand two-phase flow in hydraulic machinery and systems},
\newblock \bibinfo{journal}{Int. J. Heat Fluid Flow} \bibinfo{volume}{95}
  (\bibinfo{year}{2022}) \bibinfo{pages}{108974}.
\bibitem[{Gai et~al.(2020)Gai, Hadjadj, Kudriakov, and
  Thomine}]{gai2020particles}
\bibinfo{author}{G.~Gai}, \bibinfo{author}{A.~Hadjadj},
  \bibinfo{author}{S.~Kudriakov}, \bibinfo{author}{O.~Thomine},
\newblock \bibinfo{title}{Particles-induced turbulence: A critical review of
  physical concepts, numerical modelings and experimental investigations},
\newblock \bibinfo{journal}{Theor. Appl. Mech. Lett.} \bibinfo{volume}{10}
  (\bibinfo{year}{2020}) \bibinfo{pages}{241--248}.
\bibitem[{Wang et~al.(2019)Wang, Zhao, and Yao}]{wang2019large}
\bibinfo{author}{Y.~Wang}, \bibinfo{author}{Y.~Zhao}, \bibinfo{author}{J.~Yao},
\newblock \bibinfo{title}{Large eddy simulation of particle deposition and
  resuspension in turbulent duct flows},
\newblock \bibinfo{journal}{Adv. Powder Technol.} \bibinfo{volume}{30}
  (\bibinfo{year}{2019}) \bibinfo{pages}{656--671}.
\bibitem[{Anzoom et~al.(2024)Anzoom, Bournival, and Ata}]{anzoom2024coarse}
\bibinfo{author}{S.~J. Anzoom}, \bibinfo{author}{G.~Bournival},
  \bibinfo{author}{S.~Ata},
\newblock \bibinfo{title}{Coarse particle flotation: A review},
\newblock \bibinfo{journal}{Miner. Eng.} \bibinfo{volume}{206}
  (\bibinfo{year}{2024}) \bibinfo{pages}{108499}.
\bibitem[{Zhao et~al.(2023)Zhao, Liu, Lin, Chen, Chen, and
  Wang}]{zhao2023estimation}
\bibinfo{author}{L.~Zhao}, \bibinfo{author}{D.~Liu}, \bibinfo{author}{J.~Lin},
  \bibinfo{author}{L.~Chen}, \bibinfo{author}{S.~Chen},
  \bibinfo{author}{G.~Wang},
\newblock \bibinfo{title}{Estimation of turbulent dissipation rates and its
  implications for the particle-bubble interactions in flotation},
\newblock \bibinfo{journal}{Miner. Eng.} \bibinfo{volume}{201}
  (\bibinfo{year}{2023}) \bibinfo{pages}{108230}.
\bibitem[{Narayanan et~al.(1969)Narayanan, Bhatia, and
  Guha}]{narayanan1969suspension}
\bibinfo{author}{S.~Narayanan}, \bibinfo{author}{V.~Bhatia},
  \bibinfo{author}{D.~Guha},
\newblock \bibinfo{title}{Suspension of solids by bubble agitation},
\newblock \bibinfo{journal}{Can. J. Chem. Eng.} \bibinfo{volume}{47}
  (\bibinfo{year}{1969}) \bibinfo{pages}{360--364}.
\bibitem[{Murai(2014)}]{murai2014frictional}
\bibinfo{author}{Y.~Murai},
\newblock \bibinfo{title}{Frictional drag reduction by bubble injection},
\newblock \bibinfo{journal}{Exp. Fluids} \bibinfo{volume}{55}
  (\bibinfo{year}{2014}) \bibinfo{pages}{1--28}.
\bibitem[{Gabillet et~al.(2002)Gabillet, Colin, and
  Fabre}]{gabillet2002experimental}
\bibinfo{author}{C.~Gabillet}, \bibinfo{author}{C.~Colin},
  \bibinfo{author}{J.~Fabre},
\newblock \bibinfo{title}{Experimental study of bubble injection in a turbulent
  boundary layer},
\newblock \bibinfo{journal}{Int. J. Multiph. Flow} \bibinfo{volume}{28}
  (\bibinfo{year}{2002}) \bibinfo{pages}{553--578}.
\bibitem[{Hoque et~al.(2024)Hoque, Mitra, and Evans}]{hoque2024bubble}
\bibinfo{author}{M.~M. Hoque}, \bibinfo{author}{S.~Mitra},
  \bibinfo{author}{G.~Evans},
\newblock \bibinfo{title}{Bubble size distribution and turbulence
  characterization in a bubbly flow in the presence of surfactant},
\newblock \bibinfo{journal}{Exp. Therm. Fluid Sci} \bibinfo{volume}{155}
  (\bibinfo{year}{2024}) \bibinfo{pages}{111199}.
\bibitem[{Tao(2005)}]{tao2005role}
\bibinfo{author}{D.~Tao},
\newblock \bibinfo{title}{Role of bubble size in flotation of coarse and fine
  particles—a review},
\newblock \bibinfo{journal}{Sep. Sci. Technol.} \bibinfo{volume}{39}
  (\bibinfo{year}{2005}) \bibinfo{pages}{741--760}.
\bibitem[{Zakari et~al.(2025)Zakari, Hoque, Ireland, Evans, and
  Mitra}]{zakari2025dynamics}
\bibinfo{author}{A.~G. Zakari}, \bibinfo{author}{M.~M. Hoque},
  \bibinfo{author}{P.~Ireland}, \bibinfo{author}{G.~Evans},
  \bibinfo{author}{S.~Mitra},
\newblock \bibinfo{title}{Dynamics of gas dispersion in a rising bubble plume
  in presence of surfactant},
\newblock \bibinfo{journal}{Miner. Eng.} \bibinfo{volume}{222}
  (\bibinfo{year}{2025}) \bibinfo{pages}{109145}.

\end{thebibliography}
\end{document}